\documentclass[prd,twocolumn,reprint,preprintnumbers,nofootinbib,superscriptaddress,eqsecnum]{revtex4-1}
\usepackage{amsmath, float, graphicx,placeins,amssymb,multirow,pgfplots,pgfplotstable}
\usepackage{tikz,hyperref}
\usepackage{enumitem}
\usepackage{lipsum}
\usepackage{graphicx}
\usepackage{tabularx}
\usepackage{CJKutf8}
\usetikzlibrary{shapes.geometric,arrows}
\usepackage{amsmath,amssymb,amsthm,amsfonts}
\usepackage{graphicx,tabularx}
\usepackage{color}
\usepackage{multirow}
\usepackage{comment}
\usepackage{enumitem}
\usepackage{cleveref}
\usepackage{slashed}
\usepackage[normalem]{ulem}
\usepackage{scalerel}
\usepackage{wrapfig}
\usepackage{cancel}
\usepackage{xcolor}
\usepackage{soul}

\usepackage{lineno}

\newcommand{\abs}[1]{\left | {#1} \right|}
\usepackage{slashed}

\begin{document}

\preprint{UCI-TR-2022-16}

\title{Extending the Discovery Potential for Inelastic-Dipole Dark Matter with FASER}

\def\andname{\hspace*{-0.5em}} 
\author{Keith R.~Dienes}
\email[Email address: ]{dienes@arizona.edu}
\affiliation{Department of Physics, University of Arizona, Tucson, AZ 85721 USA}
\affiliation{Department of Physics, University of Maryland, College Park, MD 20742 USA}
\author{Jonathan L.~Feng}
\email[Email address: ]{jlf@uci.edu}
\affiliation{Department of Physics and Astronomy, University of California, Irvine, CA  92697 USA}
\author{Max Fieg}
\email[Email address: ]{mfieg@uci.edu}
\affiliation{Department of Physics and Astronomy, University of California, Irvine, CA  92697 USA}
\author{Fei Huang}
\email[Email address: ]{fei.huang@weizmann.ac.il}
\affiliation{Department of Particle Physics and Astrophysics, Weizmann Institute of Science, Rehovot 7610001, Israel}
\author{Seung J.~Lee}
\email[Email address: ]{sjjlee@korea.ac.kr}
\affiliation{Department of Physics, Korea University, Seoul 136-713, Korea}
\author{Brooks Thomas}
\email[Email address: ]{thomasbd@lafayette.edu}
\affiliation{Department of Physics, Lafayette College, Easton, PA 18042 USA}

\begin{abstract}
Neutral particles are notoriously difficult to observe through electromagnetic 
interactions.  As a result, they naturally elude detection in most collider detectors.  
In this paper, we point out that neutral particles that interact through a 
dipole interaction can nevertheless be detected in far-forward detectors 
designed to search for long-lived particles (LLPs).  In contrast to previous 
analyses that focused on neutral particles with elastic interactions, we 
consider inelastic interactions.  This naturally leads to LLPs, and we 
demonstrate that FASER (and future experiments at the Forward Physics Facility) 
will be able to probe substantial regions of the associated parameter space.  
In particular, we find that FASER is capable of probing the region of 
parameter space wherein thermal freeze-out gives rise to an 
$\mathcal{O}$(GeV)
dark-matter candidate with the appropriate relic
abundance, as well as regions of parameter space that are difficult to probe 
at fixed-target experiments.  FASER and its successor experiments may therefore 
play a critical role in the discovery of such a dark-matter candidate. 
\end{abstract}

\maketitle

\renewcommand{\baselinestretch}{0.82}\normalsize
\tableofcontents
\renewcommand{\baselinestretch}{1.0}\normalsize


\newcommand{\PRE}[1]{{#1}} 
\newcommand{\del}{\partial}
\newcommand{\nbox}{{\,\lower0.9pt\vbox{\hrule \hbox{\vrule height 0.2 cm
\hskip 0.2 cm \vrule height 0.2 cm}\hrule}\,}}

\newcommand{\postscript}[2]{\setlength{\epsfxsize}{#2\hsize}
   \centerline{\epsfbox{#1}}}
\newcommand{\gweak}{g_{\text{weak}}}
\newcommand{\mweak}{m_{\text{weak}}}
\newcommand{\mplanck}{M_{\text{Pl}}}
\newcommand{\mstar}{M_{*}}
\newcommand{\sigmaan}{\sigma_{\text{an}}}
\newcommand{\sigmatot}{\sigma_{\text{tot}}}
\newcommand{\sigmaSI}{\sigma_{\rm SI}}
\newcommand{\sigmaSD}{\sigma_{\rm SD}}
\newcommand{\OmegaM}{\Omega_{\text{M}}}
\newcommand{\OmegaDM}{\Omega_{\text{DM}}}
\newcommand{\ipb}{\text{pb}^{-1}}
\newcommand{\ifb}{\text{fb}^{-1}}
\newcommand{\iab}{\text{ab}^{-1}}
\newcommand{\ev}{\text{eV}}
\newcommand{\kev}{\text{keV}}
\newcommand{\mev}{\text{MeV}}
\newcommand{\gev}{\text{GeV}}
\newcommand{\tev}{\text{TeV}}
\newcommand{\pb}{\text{pb}}
\newcommand{\mb}{\text{mb}}
\newcommand{\cm}{\text{cm}}
\newcommand{\m}{\text{m}}
\newcommand{\km}{\text{km}}
\newcommand{\kg}{\text{kg}}
\newcommand{\g}{\text{g}}
\newcommand{\s}{\text{s}}
\newcommand{\yr}{\text{yr}}
\newcommand{\Mpc}{\text{Mpc}}
\newcommand{\etal}{{\em et al.}}
\newcommand{\eg}{{\em e.g.}}
\newcommand{\ie}{{\em i.e.}}
\newcommand{\ibid}{{\em ibid.}}
\newcommand{\eqnref}[1]{Eq.~(\ref{#1})}
\newcommand{\eqsref}[2]{Eqs.~(\ref{#1}) and (\ref{#2})}
\newcommand{\Eqnref}[1]{Equation~(\ref{#1})}
\newcommand{\expt}[1]{\langle #1 \rangle}
\newcommand{\secref}[1]{Sect.~\ref{sec:#1}}
\newcommand{\secsref}[2]{Secs.~\ref{sec:#1} and \ref{sec:#2}}
\newcommand{\Secref}[1]{Section~\ref{sec:#1}}
\newcommand{\appref}[1]{Appendix~\ref{app:#1}}
\newcommand{\figref}[1]{Fig.~\ref{fig:#1}}
\newcommand{\figsref}[2]{Figs.~\ref{fig:#1} and \ref{fig:#2}}
\newcommand{\Figref}[1]{Figure~\ref{fig:#1}}
\newcommand{\tableref}[1]{Table~\ref{table:#1}}
\newcommand{\tablesref}[2]{Tables~\ref{table:#1} and \ref{table:#2}}
\newcommand{\Dsle}[1]{\slash\hskip -0.23 cm #1}
\newcommand{\met}{{\Dsle E_T}}
\newcommand{\mpt}{\not{\! p_T}}
\newcommand{\Dslp}[1]{\slash\hskip -0.23 cm #1}
\newcommand{\Dsl}[1]{\slash\hskip -0.20 cm #1}
\newcommand{\nn}{\nonumber}
\newcommand{\mB}{m_{B^1}}
\newcommand{\mq}{m_{q^1}}
\newcommand{\mf}{m_{f^1}}
\newcommand{\mKK}{m_{KK}}
\newcommand{\WIMP}{\text{WIMP}}
\newcommand{\SWIMP}{\text{SWIMP}}
\newcommand{\NLSP}{\text{NLSP}}
\newcommand{\LSP}{\text{LSP}}
\newcommand{\mWIMP}{m_{\WIMP}}
\newcommand{\mSWIMP}{m_{\SWIMP}}
\newcommand{\mNLSP}{m_{\NLSP}}
\newcommand{\mchi}{m_{\chi}}
\newcommand{\mgravitino}{m_{\gravitino}}
\newcommand{\mmed}{M_{\text{med}}}
\newcommand{\gravitino}{\tilde{G}}
\newcommand{\Bino}{\tilde{B}}
\newcommand{\photino}{\tilde{\gamma}}
\newcommand{\stau}{\tilde{\tau}}
\newcommand{\slepton}{\tilde{l}}
\newcommand{\snu}{\tilde{\nu}}
\newcommand{\squark}{\tilde{q}}
\newcommand{\mgaugino}{M_{1/2}}
\newcommand{\epsEM}{\varepsilon_{\text{EM}}}
\newcommand{\mmess}{M_{\text{mess}}}
\newcommand{\lmess}{\Lambda}
\newcommand{\nmess}{N_{\text{m}}}
\newcommand{\signmu}{\text{sign}(\mu)}
\newcommand{\Omegachi}{\Omega_{\chi}}
\newcommand{\lambdafs}{\lambda_{\text{FS}}}
\newcommand{\be}{\begin{equation} \begin{aligned}}
\newcommand{\ee}{\end{aligned} \end{equation}}
\newcommand{\bea}{\begin{eqnarray}}
\newcommand{\eea}{\end{eqnarray}}
\newcommand{\beq}{\begin{equation}}
\newcommand{\eeq}{\end{equation}}
\newcommand{\beqn}{\begin{eqnarray}}
\newcommand{\eeqn}{\end{eqnarray}}
\newcommand{\baln}{\begin{align}}
\newcommand{\ealn}{\end{align}}
\newcommand{\lsim}{\lower.7ex\hbox{$\;\stackrel{\textstyle<}{\sim}\;$}}
\newcommand{\gsim}{\lower.7ex\hbox{$\;\stackrel{\textstyle>}{\sim}\;$}}
\newcommand{\maf}[1]{{\color {purple} [#1 - MF]}}
\newcolumntype{C}{>{\centering\arraybackslash}X}

\newcommand{\ssection}[1]{{\em #1.\ }}
\newcommand{\rem}[1]{\textbf{#1}}

\def\etc{{\it etc}.\/}
\def\calN{{\cal N}}
\def\SM{{\rm SM}}
\def\mptwo{{m_{\pi^0}^2}}
\def\mp{{m_{\pi^0}}}
\def\sqtsn{\sqrt{s_n}}
\def\sqtsn{\sqrt{s_n}}
\def\sqtsn{\sqrt{s_n}}
\def\sqts0{\sqrt{s_0}}
\def\Dsqts{\Delta(\sqrt{s})}
\def\smin{s_{\rm min}}
\def\Omegatot{\Omega_{\mathrm{tot}}}
\def\rhotot{\rho_{\mathrm{tot}}}
\def\rhocrit{\rho_{\mathrm{crit}}}
\def\OmegaDM{\Omega_{\mathrm{DM}}}
\def\OmegaDMbar{\overline{\Omega}_{\mathrm{DM}}}
\def\tLS{t_{\mathrm{LS}}}
\def\aLS{a_{\mathrm{LS}}}
\def\zLS{z_{\mathrm{LS}}}
\def\tnow{t_{\mathrm{now}}}
\def\znow{z_{\mathrm{now}}}
\def\Ndof{N_{\mathrm{d.o.f.}}\/}
\def\ra{\rightarrow}
\def\Lamcalo{\Lambda_{\mathcal{O}}}


\section{Introduction}


Since it began running almost 15 years ago, the Large Hadron Collider (LHC) 
has been one of the most effective tools we have for understanding the
Standard Model (SM) and for probing physics beyond the Standard Model (BSM). 
Among its many successes, the LHC has discovered the Higgs boson and explored its 
properties in great detail. LHC data has also hinted at new physics in the flavor 
sector, and searches involving missing energy, displaced vertices, resonances, and 
a host of other phenomena have been used to constrain the properties of potential 
dark sectors~\cite{Boveia:2018yeb,Lee:2018pag,Alimena:2019zri}. However, despite 
this progress, the LHC has not yielded any definitive discovery of BSM physics.
 
Of course, new physics might be hiding within experimental blind spots.
In particular, new physics might be found in the large-pseudorapidity
regions, where there is a large flux of SM particles.  The ForwArd Search 
ExpeRiment (FASER)~\cite{Feng:2017uoz,FASER:2018eoc,FASER:2022hcn} 
provides an ideal opportunity for probing this region.  FASER is a 
cylindrical detector placed 476~m from the ATLAS interaction point (IP) along 
the beam axis.   As such, FASER aims to capture the decay products of long-lived 
particles (LLPs). Shielded by 100~m of rock and concrete, FASER typically 
exploits the large SM activity near 
the IP as a means of dark-state production through meson decay.  The FASER 
detector is already taking data, and an upgraded detector, known as FASER2, has 
been proposed, which would enhance the sensitivity of the current detector 
during the high-luminosity LHC (HL-LHC) era.  This upgraded detector would be 
housed, along with several other experiments, within the proposed Forward Physics
Facility (FPF)~\cite{MammenAbraham:2020hex,Anchordoqui:2021ghd,Feng:2022inv}.

A common benchmark signature that FASER is well suited to probe is that of a 
dark photon decaying visibly to an $e^+e^-$ pair.  Such a signature can be used 
to constrain  potential kinetic-mixing 
couplings of magnitudes $\epsilon \sim 10^{-5} -10^{-4}$ with dark-photon masses 
$m_{A'} \lesssim 1$~GeV~\cite{Feng:2017uoz}. In addition to dark-photon
models, FASER and the FPF have discovery prospects for a variety of 
new-physics scenarios in which an LLP, such as an axion-like particle, dark scalar, or heavy 
neutral lepton, decays to a pair of detectable SM particles.  
Although such scenarios are well studied in the literature 
(see, \eg,  Ref.~\cite{FASER:2018eoc} and references therein), it is important to fully 
explore additional signatures to which FASER might also be sensitive.

In this paper, we examine the detection prospects for one such LLP signature
at FASER --- a monophoton signature that arises 
in scenarios in which the photon field couples to a pair of neutral BSM 
particles $\chi_0$ and $\chi_1$ with different masses $m_1>m_0$ via an effective 
magnetic-dipole-moment (MDM) or electric-dipole-moment (EDM) operator. 
Such a scenario is particularly interesting from a 
phenomenological perspective, in part because it arises in the context of 
inelastic-dark-matter scenarios~\cite{Han:1997wn,Hall:1997ah,Tucker-Smith:2001myb}. The inelastic MDM operator was also recently studied in the context of sterile neutrinos \cite{Barducci:2022gdv}.
If $\chi_0$ and $\chi_1$ are sufficiently light, they can be produced together 
at a significant rate at the LHC via meson-decay processes involving such 
MDM and EDM operators.  Moreover, these operators render $\chi_1$  unstable, 
since $\chi_1$ can  decay via the process 
$\chi_1\rightarrow \chi_0\gamma$. Thus, if the
resulting $\chi_1$ decay width is of the appropriate order,
particles of this species emitted in the forward direction at the LHC could 
potentially decay inside the FASER detector.
The calorimeter of the FASER detector can then detect the 
resulting monophoton signal~\cite{Kling:2022ehv,Dreiner:2022swd}.  

From a theoretical perspective, inelastic-dark-matter models of this sort 
are also of interest because they represent the simplest realization of 
a non-minimal dark sector involving $N$ dark states $\chi_i$ 
with similar quantum numbers, where $i = 0, 1, \ldots, N-1$.  The phenomenology 
of such dark sectors can differ significantly from that of 
minimal dark sectors, due to the presence of the additional dark-sector 
states (for a review, see Ref.~\cite{Dienes:2022zbh}). 
Indeed, the presence of the additional dark-sector states generically 
gives rise to decay processes in which a heavier $\chi_i$ particle decays into a 
final state involving one or more lighter $\chi_i$ particles.  Such decay
processes alter the complementarity relations between different experimental
probes of the dark sector~\cite{Dienes:2014via,Dienes:2017ylr}.  Moreover, they
can also lead to a variety of striking signatures at 
colliders~\cite{Dienes:2019krh,Dienes:2021cxr} and even open up new ways of
addressing the dark-matter problem, such as those that arise within the Dynamical 
Dark Matter (DDM) framework~\cite{Dienes:2011ja,Dienes:2011sa,Dienes:2012jb}.
Non-minimal dark sectors involving multiple states with similar quantum numbers 
--- and indeed often involving large values of $N$ --- emerge naturally in a 
variety of BSM contexts, including theories with extra spacetime dimensions, 
theories involving confining hidden-sector gauge groups, and string theory.  
A two-component inelastic-dark-matter model of the sort on which we shall focus 
in this paper may be viewed as the simplest example of a non-minimal dark 
sector in which these phenomenological possibilities arise.

One of the most important features of inelastic dark matter 
is that it provides a mechanism by which event
rates at direct-detection experiments can be suppressed.
Moreover, the dynamics involved in thermal freeze-out is significantly 
more complicated in such models, with annihilation, co-annihilation, 
up- and down-scattering, and decay processes all playing a role in 
generating the relic abundance of the lighter dark-sector state. 
Realizations of inelastic dark matter in which the dark-sector states
couple to the visible sector via dipole-moment interactions 
in particular can give rise to a dark-matter abundance in accord
with observation without running afoul of the applicable  
constraints~\cite{Chang:2010en,Izaguirre:2015zva,Masso:2009mu}.  
Indeed, data from a variety of sources can constrain the parameter space of
inelastic-dipole-dark-matter models, including data from collider experiments,
fixed-target experiments, and gamma-ray telescopes.
That said, there are also interesting regions of this parameter 
space in 
which existing searches have little or no sensitivity. Notably these include those in which the spectra are highly compressed and the lighter dark-sector state 
accounts for the entirety of the observed dark-matter abundance. 

In this paper, we investigate the extent to which searches for a monophoton 
signature of $\chi_1\rightarrow\chi_0\gamma$ decays at FASER and FASER2 can 
constrain the parameter space of such inelastic-dipole-dark-matter models. 
As we shall see, these detectors provide coverage of this 
parameter space within certain regions that are not well
constrained by existing searches.  Indeed, FASER and FASER2 
can probe parameter-space regions within which the splitting between the masses 
$m_1$ and $m_0$ of the two dark-sector states $\chi_1$ and $\chi_0$ is roughly 
$\Delta m \equiv m_1 - m_0 \sim \mathcal{O}(10^{-2}) m_0$ and where
$m_0 \lesssim 5$~GeV.~  Remarkably, the reach of FASER and FASER2 within that
parameter space includes regions wherein the relic abundance of $\chi_0$ 
obtained from thermal freeze-out agrees with observed abundance of dark matter.
We note also that the small mass-splittings $\Delta m$ that make the heavier state 
$\chi_1$ long-lived also suppress the energy of the photon into which it decays,
thereby rendering the resulting monophoton signal challenging to detect at 
fixed-target experiments.  At FASER, however, this suppression is compensated by  
the significant boosts provided to these decaying particles as a result of 
the high collision energies achieved at the LHC.  Thus, as we shall see, FASER is 
capable of probing regions of parameter space that are difficult to probe at 
fixed-target experiments.  

This paper is organized as follows. In \secref{DIDM}, 
we review the manner in which the dark-sector states couple to the 
visible sector in inelastic-dipole-dark-matter models. 
In \secref{Signal at FASER}, we calculate the event rates for the signal 
and background processes relevant for monophoton searches at FASER and
FASER2 within the context of these models and assess the discovery 
reach of these detectors within the parameter space of these models.
In \secref{RelicAbundance}, we examine the annihilation
and co-annihilation processes involved in the thermal freeze-out of $\chi_0$
and $\chi_1$ in the early universe and identify the regions of parameter 
space within which relic $\chi_0$ particles constitute the majority of the dark 
matter at the present time.  In \secref{ExistingConstraints}, we examine the
additional considerations that constrain the parameter space of
inelastic-dipole-dark-matter models in the region where FASER is sensitive.
In \secref{Conclusion}, we conclude by summarizing our results and 
discussing possible directions for future work.  Details of the calculation 
of meson branching fractions and thermal relic densities, along with a brief 
discussion of the detection prospects at SHiP~\cite{Bonivento:2013jag,SHiP:2015vad}, 
a proposed experiment complementary to FASER which will probe comparable lifetimes, are given in the Appendices.


\section{Inelastic-Dipole Dark Matter\label{sec:DIDM}}


We consider an inelastic-dark-matter model in which the two dark-sector particles
$\chi_0$ and $\chi_1$ are fermions that couple to the visible sector via an 
effective MDM or EDM operator in the interaction Lagrangian of the form
\begin{eqnarray}
  {\cal{O}}_m ~&=&~ \frac{1}{\Lambda_m} 
    \overline{\chi}_1\sigma^{\mu\nu}\chi_0 F_{\mu \nu} + \rm{h.c.}
    \nonumber \\
  {\cal{O}}_e ~&=&~ \frac{1}{\Lambda_e} 
    \overline{\chi}_1\sigma^{\mu\nu}\gamma^5\chi_0 F_{\mu \nu} + \rm{h.c.}~,
    \label{eq:lagrangianEM}
\end{eqnarray}
where $F_{\mu \nu}$ is the field-strength tensor for the photon field and where
$\Lamcalo$, with $\mathcal{O} = \{m,e\}$, denotes the corresponding 
operator-suppression scale\footnote{An alternative 
parametrization is often used in the literature in which the strengths of the 
effective MDM and EDM interactions are parametrized by the dimensionful coupling 
coefficients $\mu_\chi$ and $d_\chi$, which are related of our 
operator-suppression scales $\Lambda_m$ and $\Lambda_e$ by 
$\mu_\chi = 2/\Lambda_m$ and $d_\chi = 2/\Lambda_e$.}.
We note that such effective operators can arise in a variety of UV-physics 
scenarios.  For example, they can arise as a consequence of loop-level 
processes involving heavy particles with masses of 
$\mathcal{O}(\mathrm{TeV})$ or higher that carry SM charges. 
At scales above the electroweak-symmetry-breaking scale, these operators 
would presumably be replaced by operators with  a similar structure,
but with the $U(1)_Y$ and/or $SU(2)_L$ gauge bosons present in the 
unbroken phase of the SM in place of the photon \cite{Izaguirre:2015zva}.  
However, this 
phase of the SM is not relevant for the phenomenological implications of 
inelastic-dipole dark matter that we consider in this paper.

We note that these operators can also be viewed as arising from terms 
in the elastic tensor and pseudo-tensor currents 
$J^{\mu\nu}=\overline{\chi} \sigma^{\mu\nu}\chi$ 
and $J^{\mu\nu 5}=\overline{\chi} \sigma^{\mu\nu}\gamma^5\chi$ involving 
a Dirac fermion $\chi$ that also acquires a Majorana 
mass~\cite{Izaguirre:2015zva,Chang:2010en,Masso:2009mu}.  Such a Majorana 
mass gives rise to a mass-splitting between the fermion 
mass-eigenstates --- states that play the role of our $\chi_0$ and 
$\chi_1$ --- and to operators with the inelastic coupling structures 
given in \eqnref{eq:lagrangianEM}.  We parametrize this mass-splitting in 
terms of the dimensionless ratio
\begin{equation}
    \Delta ~\equiv~ \frac{\Delta m}{ m_0} ~\equiv~ \frac{m_1-m_0}{m_0}~. 
\end{equation}
Since Dirac and Majorana terms in the dark-fermion mass matrix need not 
necessarily have a common origin, $\Delta$ can in principle take a broad 
range of values.  As such, we take the parameter space of our 
inelastic-dipole-dark-matter model to be characterized by the three
parameters $m_0$, $\Delta$, and $\Lamcalo$. 

The interactions in \eqnref{eq:lagrangianEM} have a variety of 
phenomenological implications.  For example, in the presence of such
interactions, $\chi_0$ and $\chi_1$ particles can be produced both at 
colliders and at beam-dump experiments via processes of the form 
$f \bar{f}\rightarrow \chi_0 \overline{\chi}_1$, where $f$ denotes a SM 
fermion, via bremsstrahlung-like processes involving a virtual photon, 
or via the decays of vector and pseudoscalar mesons.  Once produced,  
these particles can be detected in  a variety of ways due to the 
up-scattering, down-scattering, and decay processes to which 
these same interactions give rise.  Indeed, potential experimental 
signatures of inelastic-dipole dark matter include the scattering of 
$\chi_0$ or $\chi_1$ with atomic nuclei, events involving sizable missing 
transverse energy at colliders, and rare meson 
decays~\cite{Chu:2018qrm,Kling:2022ykt,Chu:2020ysb}.  Existing limits
on these processes place non-trivial constraints on our parameter space. 

The decay process $\chi_1\rightarrow \chi_0 \gamma$ provides an
additional experimental probe of inelastic-dipole-dark-matter models.
Indeed, searches for decays of this sort at $B$-factories and at the main LHC 
detectors likewise place non-trivial constraints on our parameter
space~\cite{Izaguirre:2015zva}.  
However, as discussed in the Introduction, this same decay process can also 
potentially lead to signatures at FASER and FASER2 within regions of that 
parameter space that are largely unconstrained.  
One particularly interesting such region of parameter space turns out to be 
that in which $\Delta\sim{\cal{O}}(0.01)$ and
$\mathcal{O}({\rm MeV}) < m_0 < \mathcal{O}({\rm GeV})$.  Within this 
region, not only can a significant flux of $\chi_i$ pairs be generated in 
the forward direction, but the lifetime $\tau$ of $\chi_1$ is also such that 
a significant fraction of $\chi_1$ decays occur within the 
FASER detector.  Indeed, as we shall justify below, the lab-frame decay length 
$\bar{d}=\gamma\beta\tau$, where $\beta$ and $\gamma$ are the usual relativistic 
factors, can be expressed as a product of factors of the form
\begin{eqnarray}
\bar{d} &~\approx~&
  (600{\,\rm{m}})\times
  \left(\frac{500~ {\rm{MeV}}}{m_0}\right)^4
  \left(\frac{0.01}{\Delta}\right)^3 \nonumber \\ & & \times
  \left(\frac{\Lamcalo}{15.5\,\rm TeV}\right)^2
  \left(\frac{E_{\chi_1}}{{\rm TeV}}\right)~,
  \label{dbarparam}
\end{eqnarray}
where $E_{\chi_1}$ is the energy of $\chi_1$.
Given that FASER is located approximately $476{\,\rm{m}}$ 
away from the ATLAS detector, this expression for $\bar{d}$ 
makes it clear that the region of parameter space within which 
$\Delta\sim{\cal{O}}(0.01)$ and 
$\mathcal{O}({\rm MeV}) < m_0 < \mathcal{O}({\rm GeV})$,
while $\Lamcalo$ is near the TeV scale, 
is one in which the prospects of observing $\chi_1$ decays insider 
FASER are particularly auspicious.  Since the proposed location
of FASER2 within the FPF lies a similar distance away from the ATLAS 
detector, this region of parameter space is auspicious for 
detection at FASER2 as well.

As we shall see, this region of inelastic-dipole-dark-matter parameter 
space is interesting not only because it is auspicious for detection at 
FASER; it is also interesting from a dark-matter perspective.  
Indeed, within this region the
abundance of $\chi_0$ obtained from thermal freeze-out and the 
subsequent decays of $\chi_1$ can accord with the dark-matter 
abundance inferred from Planck data~\cite{Planck:2018vyg}.
Moreover, since the value of the mass-splitting 
$\Delta m \gtrsim 10\,\mathrm{keV}$ 
within this parameter-space region of interest lies well above the lab-frame 
kinetic energy of a typical $\chi_0$ particle of mass 
$\mathcal{O}({\rm MeV}) < m_0 < \mathcal{O}({\rm GeV})$ in the Milky-Way halo, 
event rates at direct-detection experiments are highly suppressed.
As a result, such experiments are insensitive to dipole-interacting 
dark matter within this region of parameter space.


\section{Signal at FASER\label{sec:Signal at FASER}}


\subsection{Production\label{sec:chiproduction}}

A number of processes contribute to the overall production rate for 
$\chi_1$ particles in the far-forward direction at the LHC in 
inelastic-dipole-dark-matter models.  These include production from meson 
decay, Drell-Yan production, and secondary 
production --- \ie, the production of $\chi_1$ particles due to the 
up-scattering of $\chi_0$ particles produced at the IP as they travel 
through the intervening material between the IP and the FASER 
detector~\cite{Jodlowski:2019ycu}.

The decays of neutral mesons produced at the ATLAS IP turn out to provide the dominant contribution to 
this overall production rate.  Indeed both vector and pseudoscalar mesons are 
produced at the LHC in copious amounts in the forward direction towards the FASER detector, which is coaxial with the ATLAS experiment.  Moreover, the decays of both kinds of 
meson contribute to $\chi_1$ production.  Within the regime in which
$\mathcal{O}({\rm MeV}) \lesssim (2+\Delta) m_0 \lesssim \mathcal{O}({\rm GeV})$,
the relevant pseudoscalar mesons are $\pi^0$, $\eta$, and $\eta'$, while the relevant 
vector mesons are $\rho$, $\omega$, $\phi$, $J/\psi$, $\psi(2S)$, and $\Upsilon(nS)$.
The dominant contribution to $\chi_1$ production from each of these meson 
species $M$ is due to the decays of mesons that are produced in the 
far-forward direction --- \ie, with pseudorapidities $\eta > \eta_F$, 
where $\eta_F = 9.2$ and $\eta_F = 7.2$ for FASER and FASER2, respectively
--- and have lab-frame energies $E_M \sim \mathcal{O}({\rm TeV})$.  Mesons with such 
energies are sufficiently highly boosted that the vast majority of $\chi_1$ particles 
produced by their decays will likewise travel in the far-forward direction, regardless of 
the angle that the momentum vector of the $\chi_1$ makes with the momentum 
vector of the decaying meson in the rest frame of the meson.  Thus, we may 
obtain a rough estimate of the total cross-section
$\sigma_{\chi_1 M}^{({\rm eff})}$ for the production of $\chi_1$ particles 
in the direction of FASER from the decay of a particular meson species $M$ 
simply by multiplying the production cross-section $\sigma_M$ for 
mesons of that species with $\eta \geq \eta_F$
by the overall branching fraction for $M$ into final states including $\chi_1$.
We obtain the $\sigma_M$ for each relevant meson species from the {\tt FORESEE} 
code package~\cite{Kling:2021fwx}.

A pseudoscalar meson $P$ produces $\chi_i$ particles primarily through the three-body decay process $P\rightarrow \gamma\gamma^*\rightarrow\overline{\chi}_0\chi_1\gamma$. One could also consider the two-body decay $P \rightarrow \bar{\chi}_0 \chi_1$, but the branching fraction for this process is loop-suppressed (because the leading diagram arises at one-loop order), helicity-suppressed, and further suppressed by $\Lamcalo^{-4}\sim (10\,{\rm TeV})^{-4}$, so we shall not consider this production process further.
By contrast, a vector meson $V$ produces these particles primarily through the two-body decay 
process $V\rightarrow\overline{\chi}_0\chi_1$.  The corresponding decay processes for the 
case of an elastic MDM or EDM interaction were considered in  
Refs.~\cite{Chu:2018qrm,Chu:2020ysb,Kling:2022ykt}. 
Full expressions for the branching fractions 
${\rm BR}(V\rightarrow\overline{\chi}_0\chi_1)$ and
${\rm BR}(P\rightarrow \overline{\chi}_0\chi_1\gamma)$ for the aforementioned 
vector- and pseudoscalar-meson decay processes are provided in 
Appendix~\ref{app:BFmesons} for both the MDM- and an EDM-interaction cases.  
We note that both ${\rm BR}(V\rightarrow\overline{\chi}_0\chi_1)$ and
${\rm BR}(P\rightarrow \overline{\chi}_0\chi_1\gamma)$ are, to a good approximation, 
proportional to the square of the meson mass $m_M$ in the $(2 + \Delta)m_0 \ll m_M$
regime.  Thus, although the cross-sections for the production of light pseudoscalars 
such as the $\pi^0$ in the forward direction at the LHC are significantly higher than 
those for much heavier vector mesons, the likelihood that each such vector meson 
will produce a $\chi_1$ when it decays is enhanced simply by virtue of its 
being heavier.  Moreover, the
${\rm BR}(P\rightarrow \overline{\chi}_0\chi_1\gamma)$ for all pseudoscalar mesons,
regardless of their mass, experience an additional suppression relative to the 
${\rm BR}(V\rightarrow\overline{\chi}_0\chi_1)$ for vector mesons 
due to phase-space considerations. 
As a consequence of these two effects, we find that the decays of vector mesons 
in fact provide the dominant contribution to the production rate of $\chi_1$ 
particles in the forward direction at the LHC within our parameter-space region of 
interest.   

Given that vector-meson decays to $\chi_i$ pairs dominate the production
rate of $\chi_1$ particles in the forward direction, it is instructive to 
compare the branching-fraction expressions
${\rm BR}_{\rm MDM}(V\rightarrow\overline{\chi}_0\chi_1)$ and 
${\rm BR}_{\rm EDM}(V\rightarrow\overline{\chi}_0\chi_1)$ for
these decays in the MDM- and EDM-interaction cases, respectively.
In the regime in which $m_V \gg (2+\Delta) m_0$, these two expressions 
coincide and are well approximated by the single expression
\begin{equation}
    {\rm BR}(V\rightarrow\overline{\chi}_0\chi_1) ~\approx~
      \frac{m_V^2}{8\pi\alpha\Lamcalo^2}\,{\rm BR}(V\rightarrow e^+e^-)~,
    \label{VM branching fraction}
\end{equation}

where $\alpha$ is the fine-structure constant. By contrast, as the dark-fermion masses are increased to the point at which 
$(2+\Delta)m_0$ becomes comparable with $m_V$, the expressions for
${\rm BR}_{\rm MDM}(V\rightarrow\overline{\chi}_0\chi_1)$ and
${\rm BR}_{\rm EDM}(V\rightarrow\overline{\chi}_0\chi_1)$ begin to deviate
from each other appreciably.  For example, in the elastic limit --- \ie, 
the limit in which $\Delta \rightarrow 0$ --- the ratio of these two 
branching-fraction expressions takes the particularly simple form
\be
  \frac{{\rm BR}_{\rm MDM}(V\rightarrow\overline{\chi}_0\chi_1)}
    {{\rm BR}_{\rm EDM}(V\rightarrow\overline{\chi}_0\chi_1)}
    ~\xrightarrow{\Delta\rightarrow 0}~
    \frac{\Lambda_e^2}{\Lambda_m^2}
    \left(\frac{M_V^2+8m_{0}^2}{M_V^2-4m_{0}^2}\right)~.~
\ee
Thus, in this limit, the production rate of $\chi_i$ pairs from vector-meson decay
is larger for an MDM interaction than it is for an EDM interaction with the same value 
of $\Lamcalo$.  We find that this qualitative result likewise holds in the inelastic
case in which $\Delta \neq 0$ as well.  This can be understood as a consequence of 
angular-momentum conservation.  In the decaying meson's center-of-mass frame, 
the MDM operator creates the $\overline{\chi}_0\chi_1$ final-state in the 
$| S \, S_z \rangle = | 1 \, 0 \rangle$ spin state, and since angular-momentum 
conservation requires $J=1$, the final-state orbital angular momentum is $L=0$.  
By contrast, the EDM operator creates a final state with spin state 
$| S \, S_z \rangle = | 0 \, 0 \rangle$ and $L=1$. As a result, the partial 
width in the EDM case is $P$-wave suppressed, as can be seen in the squared amplitude 
in Eq.~(\ref{eq:EDMamplitude}), and thus FASER will have greater sensitivity 
for large $m_0$ for the MDM case.

\begin{figure*}
    \centering
    \includegraphics[scale=0.45]{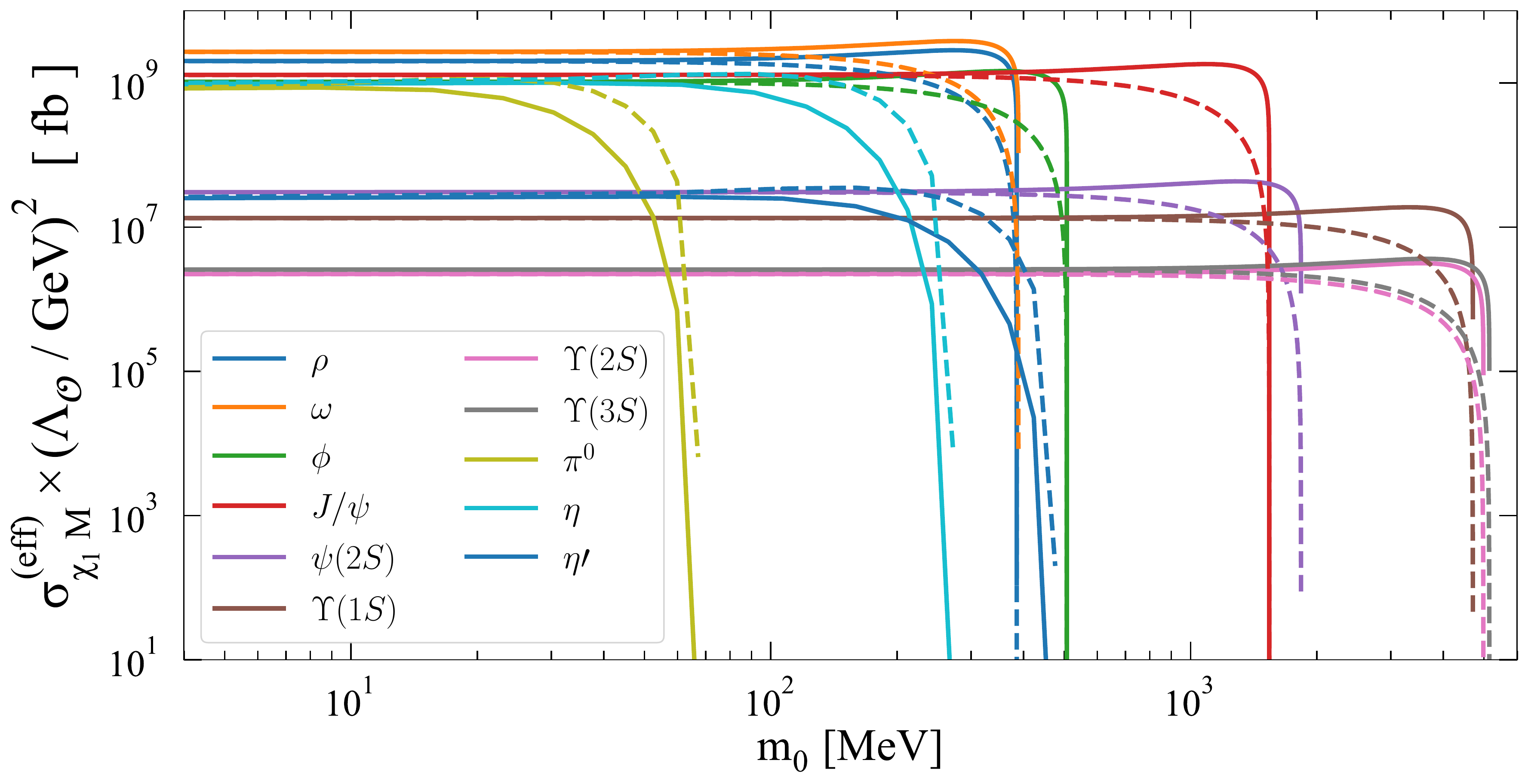}
    \caption{Effective cross-sections $\sigma_{\chi_1 M}^{({\rm eff})}$
    for the production of $\chi_1$ particles in the direction of FASER from the decay of 
    several relevant meson species $M$, shown as functions of $m_0$ for $\eta_F = 9.2$.
    In order to display our results in as model-independent a manner as possible, we 
    scale each cross-section by a factor of $(\Lamcalo/{\rm GeV})^2$ 
    to cancel the overall factor of $\Lamcalo^{-2}$ common to all of the 
    $\sigma_{\chi_1 M}^{({\rm eff})}$.  The solid curves correspond to the case 
    of a MDM operator, while the dashed curves correspond to the case of an EDM
    operator.  The results shown here correspond to the case in which $\Delta = 0.01$; 
    however, we note that these results are not particularly sensitive to the value of 
    $\Delta$, provided that $\Delta \ll 1$.
    \label{fig:spectra}}
\end{figure*}

In \figref{spectra}, we show the effective cross-sections $\sigma_{\chi_1 M}^{({\rm eff})}$
for the production of $\chi_1$ particles in the direction of FASER from the decay of several
relevant meson species $M$ as functions of $m_0$ for the choice of $\Delta = 0.01$.
In order to display our results in as model-independent a manner as possible, we 
scale each effective cross-section by a factor of $(\Lamcalo/{\rm GeV})^2$ to 
cancel the overall factor of $\Lamcalo^{-2}$ common to all of the 
$\sigma_{\chi_1 M}^{({\rm eff})}$.  We observe that for this value of $\Delta$,
the decays of vector mesons collectively provide the dominant contribution to the 
production rate of $\chi_1$ particles in the direction of FASER over the entire
range of $m_0$ shown.
Indeed, even for values of $m_0$ sufficiently small that the decay process 
$\pi^0 \rightarrow \gamma \overline{\chi}_0\chi_1$ is kinematically 
accessible, we find that the contribution to the total production rate from 
$\pi^0$ decay is subleading in comparison with the contributions from the 
decays of the heavier vector mesons $\omega$ and $\phi$.  Moreover, we note that 
these results are not particularly sensitive to the value of $\Delta$, provided 
that $\Delta \ll 1$.

In addition to the contribution from meson decays, a contribution to the
overall production rate of $\chi_1$ particles at the LHC also arises from 
Drell-Yan processes of the form $q \bar{q} \rightarrow \overline{\chi}_1\chi_0$.  
Since kinematic considerations imply that only mesons $M$ with masses 
$m_M > (2+\Delta) m_0 $ contribute to the production of $\chi_1$ 
particles through their decays, this Drell-Yan contribution can in principle 
be important in the regime in which $(2+\Delta)m_0$ is large.  In order to 
determine the size of this latter contribution, we evaluate the differential 
cross-section $d^2\sigma/(dE_{\chi_1}d\eta_{\chi_1})$ for the Drell-Yan production of 
$\chi_1$ particles as a function of their energy $E_{\chi_1}$ and 
pseudorapidity $\eta_{\chi_1}$ using the {\tt MG5\_aMC@NLO}~\cite{Alwall:2014hca} 
code package.  We focus on the regime in which $(2+\Delta) m_0 > 1$~GeV, as
parton-distribution functions are not well defined for momentum transfers 
below this scale.  Based on the results of these simulations, we find that the 
contribution to the production rate of $\chi_1$ particles from Drell-Yan
processes is subleading in comparison with the contribution from 
meson decay.

Yet another contribution to the production rate for $\chi_1$ arises due to
secondary production~\cite{Jodlowski:2019ycu}. Using {\tt MG5\_aMC@NLO} we evaluate this contribution by calculating the phase-space distribution of $\chi_1$ after up-scattering via $\chi_0 + A \rightarrow \chi_1 + A$ off a nucleus $A$ in the material located between the ATLAS IP and FASER. We incorporate appropriate form factors for the concrete, rock  and for the FASER$\nu$ tungsten target and calculate the event rate for subsequent $\chi_1$ decay in FASER~\cite{Jodlowski:2019ycu}. The region of parameter space for which this effect becomes potentially important is that within which $10^1~{\rm GeV} \lesssim \Lamcalo \lesssim 10^3$ GeV and $m_0 \sim {\cal O}(100)$ MeV. 
In particular we find that within this region of parameter space secondary production off tungsten in FASER$\nu$ yields a potentially significant contribution to the signal rate. By contrast, we find that secondary production off the concrete and rock located between the ATLAS IP and FASER yields a subleading contribution to the signal rate. 

While secondary production can indeed be important within this region of parameter space, this region turns out to be excluded by data from BaBar, CHARM-II, and LEP, as we shall discuss in detail in Sect.~\ref{sec:ExistingConstraints}. By contrast within regions of parameter space not excluded by current data, we find that the contribution to the total event rate is typically well below $1\%$ of the contribution from meson decay.

\subsection{Signal Rate\label{sec:signalrateatfaser}}

After a $\chi_1$ particle is produced at ATLAS in our inelastic-dipole-dark-matter 
model, it decays via the process $\chi_1\rightarrow \chi_0 \gamma$.  The
lab-frame energy of the photon produced by the decay of a $\chi_1$ particle 
with lab-frame energy $E_{\chi_1}$ is
\begin{equation}
  E_\gamma ~=~ \frac{\Delta (2+\Delta)}{2(1+\Delta)} 
     \Big[\gamma + (\gamma^2 -1)^{1/2}\cos\theta_{\gamma\chi_1}\Big] m_0~,    
\end{equation}
where $\gamma \equiv E_{\chi_1}/m_1$ is the usual relativistic boost factor 
and where $\theta_{\gamma\chi_1}$ is the angle in the rest frame of the $\chi_1$ 
particle between the momentum vector of the photon and the axis along which the 
$\chi_1$ particle is traveling.  Since the $\chi_1$ particle effectively inherits
its boost factor from the decaying meson that produced it, and since the 
characteristic energy scale for the mesons is $E_M \sim \mathcal{O}(1~{\rm TeV})$, 
the characteristic size for this boost factor is 
$\gamma \sim E_M/m_M \sim \mathcal{O}(10^3)$.  Thus, the characteristic photon 
energy for these decays within our parameter space is 
$E_\gamma \approx \gamma m_0 \Delta \sim \mathcal{O}(10~{\rm GeV})$.  
A single photon with such an energy can be detected by the calorimeter 
in either FASER or FASER2~\cite{FASER:2022hcn}.   

The probability of observing a 
statistically significant number of monophoton events at FASER during Run~3 or 
at FASER2 during the HL-LHC era depends both on the geometry of the corresponding 
detector and on the decay properties of $\chi_1$ itself.
FASER and FASER2 each have a cylindrical geometry in which the cylinder is coaxial
with the beam-collision axis at the ATLAS IP.  The components of FASER and 
FASER2 each include, starting from the side closest to the ATLAS detector, 
a dedicated decay volume, a tracking volume interleaved with a series of 
charged-particle trackers, a ``pre-shower'' station, and a calorimeter.  
Further details about each component are provided in Ref.~\cite{FASER:2022hcn}.
The radius $R$ of the cylindrical decay and tracking volumes, their combined length $h$ 
along the axial direction of the cylinder, the distance $L$ between the side of 
the decay volume closest to the ATLAS detector and the ATLAS IP, 
the value of $\eta_F$ that follows from the values of $L$ and $h$, and the integrated 
luminosity $\mathcal{L}_{\rm int}$ expected at ATLAS during the corresponding 
running period are provided in \tableref{Fasergeometry}.

We note that the values of $h$ given in \tableref{Fasergeometry} include both
the dedicated decay volume of the detector and the tracking volume.  Indeed,
the detection of a single photon within the calorimeter of FASER or FASER2 
does not rely on information from the tracker.  Moreover, the scintillators 
separating the decay volume and tracking chambers of either detector are largely 
transparent to photons with $E_\gamma \sim \mathcal{O}(10~{\rm GeV})$ energies.
Taken together, these two considerations imply that the tracking volume can also 
be viewed as an extension of the decay volume in searches for a monophoton signal 
of $\chi_1\rightarrow \chi_0 \gamma$ decays within these detectors. 

\begin{table}[h]
\begin{tabular}{||c||c|c|c|c|c||}
\hline
\hline
Detector & $h$ & $R$ & $L$  & $\eta_F$ & ${\cal{L}}_{\rm int}$ \\ \hline\hline
FASER & \multicolumn{1}{c|}{~3.5 m~~}    & \multicolumn{1}{c|}{~10 cm~~} 
& 476 m  & \textcolor{black}{9.2} & ~300 fb$^{-1}$~~ \\ \hline
~FASER2~~ & 20 m & 1 m & ~620 m~ &  \textcolor{black}{7.2} & 3 ab$^{-1}$    \\ \hline\hline
\end{tabular}
\caption{The parameters that characterize the detector geometry of FASER and FASER2, 
  along with the integrated luminosity expected at ATLAS during the corresponding running 
  period for each detector. }
\label{table:Fasergeometry}
\end{table}

The characteristic decay length of $\chi_1$ in the lab frame is given by
\begin{equation}
  \bar{d} ~=~ \frac{\abs{\vec{\mathbf{p}}_{\chi_1}}}{m_1 \Gamma_{\chi_1}}~,
  \label{eq:Shakespeare}
\end{equation}
where $\vec{\mathbf{p}}_{\chi_1}$ and $\Gamma_{\chi_1}$ denote the lab-frame momentum and 
proper decay width of $\chi_1$, respectively.  Since all of the meson species that
contribute significantly to the production rate of $\chi_1$ particles decay promptly
within the ATLAS detector, $\bar{d}$ represents the characteristic distance away from 
the IP that $\chi_1$ particles travel before they decay.
Within our parameter-space regime the proper decay width of $\chi_1$ is well 
approximated by 
\be
  \Gamma_{\chi_1} ~=~ \frac{1}{2\pi\Lamcalo^2}\left(\frac{m_1^2-m_0^2}{m_1}\right)^3
     ~\approx~\frac{4m_0^3\Delta^3}{\pi\Lamcalo^2}~.
  \label{eq:GammaChi1}
\ee
in both the MDM and EDM cases.

Since $R\ll h$ for both FASER and FASER2, we may approximate the 
distance that a $\chi_1$ particle would traverse inside the decay volume of
the detector if it didn't decay as $h$, regardless of its pseudorapidity 
$\eta_{\chi_1}$.  Thus, the probability that a given $\chi_1$ will decay 
inside FASER or FASER2 is approximately 
\begin{equation}
  P(\bar{d},h,L) ~\approx~ e^{-L/\bar{d}}(1-e^{-h/\bar{d}})~.
  \label{eq:PL}
\end{equation}
The total expected number of signal events from $\chi_1$ decays within the FASER or
FASER2 detector can therefore be approximated as
\begin{eqnarray}
  N_{\chi_1} &~\approx~& \mathcal{L}_{\rm int}\int_{\eta_F}^\infty d\eta_{\chi_1}
    \int_{m_1}^\infty dE_{\chi_1} \Bigg[\sum_M
  \frac{d\sigma^{({\rm eff})}_{\chi_1 M}}{dE_{\chi_1}d\eta _{\chi_1} } \nonumber \\
    && ~~~~~~~~~~~~~~~~~~~~ \times
    P\Big(\bar{d}(E_{\chi_1}),h,L)\Big)\Bigg]~,
\end{eqnarray}
where $d\sigma^{({\rm eff})}_{\chi_1 M}/(dE_{\chi_1}d\eta _{\chi_1})$ is the 
differential cross-section for the production of $\chi_1$ particles as a
a function of $E_{\chi_1}$ and $\eta_{\chi_1}$.  We evaluate each of these
differential cross-sections numerically by integrating the product 
of the differential production cross-section for $M$ obtained from
the \texttt{FORESEE} package~\cite{Kling:2021fwx} and the appropriate 
phase-space factor over possible meson momenta.

\subsection{Background\label{sec:background}}

The primary detector backgrounds for an $E_\gamma \sim \mathcal{O}({\rm GeV})$ 
monophoton signal of inelastic-dipole dark matter at FASER or FASER2 are those
associated with muons or neutrinos produced at the ATLAS IP --- or with muons 
produced by cosmic-ray showers --- interacting with the calorimeter 
material~\cite{Jodlowski:2020vhr,Dreiner:2022swd,Kling:2022ehv}.

Two scintillator stations located in front of the decay volume of FASER provide a 
veto on muons and other charged particles emanating from the direction of 
the ATLAS IP, the efficiency of which is around 99.99\% for each station 
individually~\cite{FASER:2022hcn}.  As a result, the contribution to the 
overall background-event rate from such muons is expected to be negligible.
An additional contribution to this overall rate also arises from events in which a muon
is produced at the IP with a pseudorapidity $\eta$ sufficiently large that it travels
in the general direction of FASER, but sufficiently small that its path does not 
intersect either of the scintillator stations.  If such a muon interacts with the 
external detector apparatus itself or some part of the surrounding infrastructure, 
this interaction can nevertheless produce a shower of particles.  Depending on the 
directions in which these particles are emitted, some of these particles could 
potentially reach and deposit their energies in the calorimeter~\cite{Arakawa:2022rmp}.
However, this contribution is also expected to be negligible, given 
that the particles produced in the shower are typically emitted in a narrow cone 
around the axis along which the muon was originally traveling.

The majority of cosmic-ray muons that pass through the FASER detector would 
likewise evade the scintillator veto.  However, these muons also behave as minimum ionizing 
particles and typically give rise to energy deposits in the FASER calorimeter of 
$\mathcal{O}({\rm MeV})$ --- an energy scale far below the energy scale associated with
the signal process.  Moreover, the contribution to the background-event rate from 
cosmic-ray muons can be further reduced by correlating timing information for calorimeter
events at FASER with timing information for collision events at ATLAS.  Given these
considerations, this contribution is also expected to be negligible. 

Finally, neutrinos produced in the far-forward direction at the ATLAS IP that 
undergo deep inelastic scattering with the material in the FASER calorimeter
can give rise to energy deposits similar to those associated with the signal process.
Background events of this sort can be distinguished from signal events
on the basis of information from the high-granularity 
preshower detector~\cite{Jodlowski:2020vhr,FASER:2018bac}.
Nevertheless, a residual contribution to the background-event rate still arises 
from two classes of neutrino events.  The first class consists of events in 
which the scattering of the neutrino with the calorimeter generates a backsplash 
involving one or more photons that are subsequently detected by the pre-shower detector.
We can significantly reduce the contribution to the background-event rate from 
the first class of events by imposing a maximum cut on the total energy deposited 
in the calorimeter, which is far higher --- typically around 
$\mathcal{O}(100~{\rm GeV})$~\cite{FASER:2019dxq} --- for these neutrino-background 
events than it is for the events that constitute our signal. 
While we do not impose such a cut explicitly in our analysis, we note that the rate of 
these backsplash photon events is not expected to be high, even at energies above 
$\mathcal{O}(100 {\rm GeV})$.  Thus, we do not expect the 
imposition of such a cut to have a significant impact on our results.
 
The second class of background events consists of those in which a neutrino 
scatters in or near the final tracking layer and produces charged particles that 
are not detected by any of the trackers, but are nevertheless detected by the 
calorimeter.  While a precise determination of this background would 
require a full detector simulation, the event rate for neutrino events of this 
class is expected to be extremely small.
Nevertheless, a TeV-neutrino scattering event is expected to produce 
${\cal O}(10)$ charged particles with larger energies than our photon signal.
Event-selection criteria designed to reject events which contain a significant
number of such particles can therefore be effective in reducing this background.

\begin{figure*}
\centering
  \includegraphics[width=0.32\textwidth]{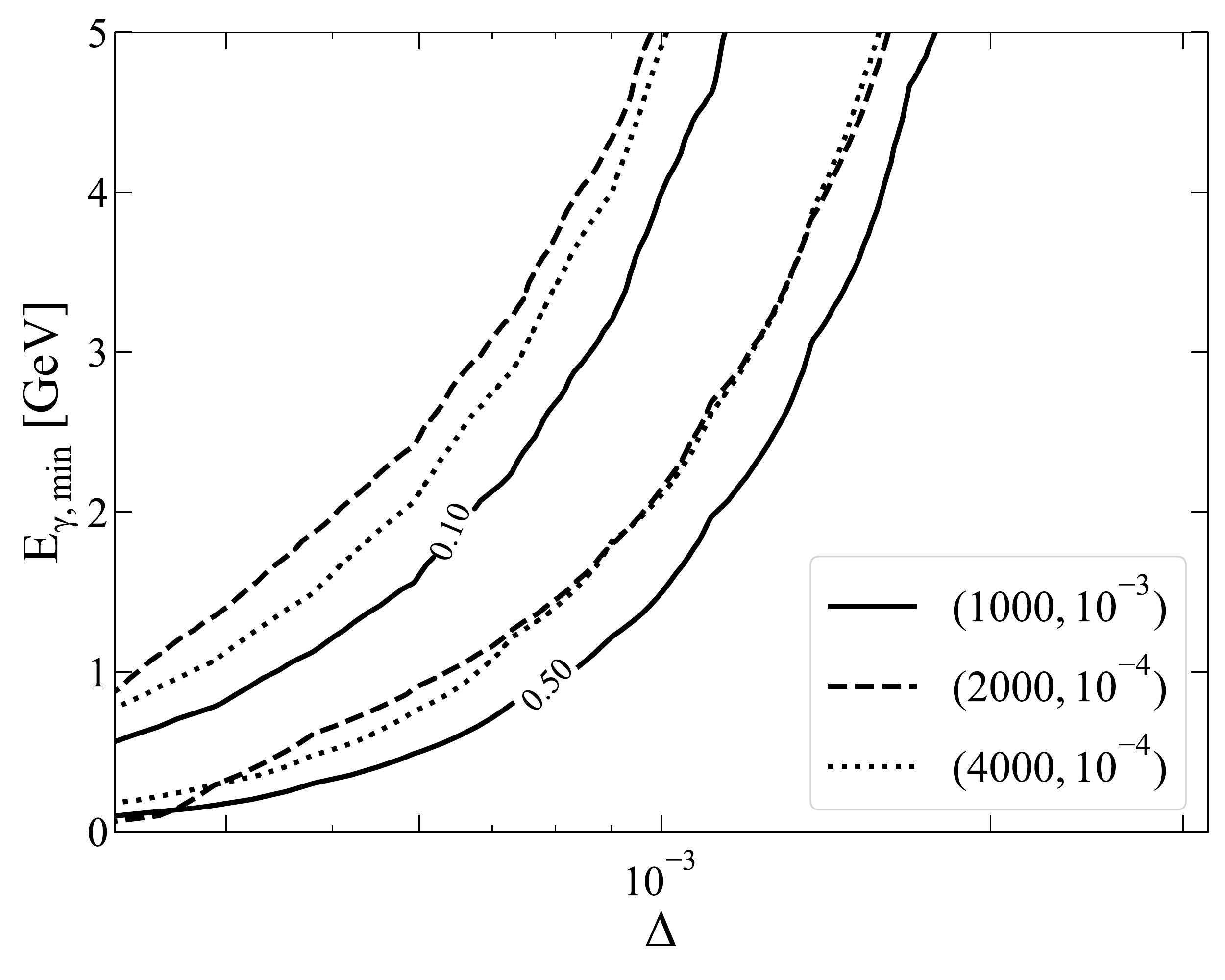}
  \includegraphics[width=0.32\textwidth]{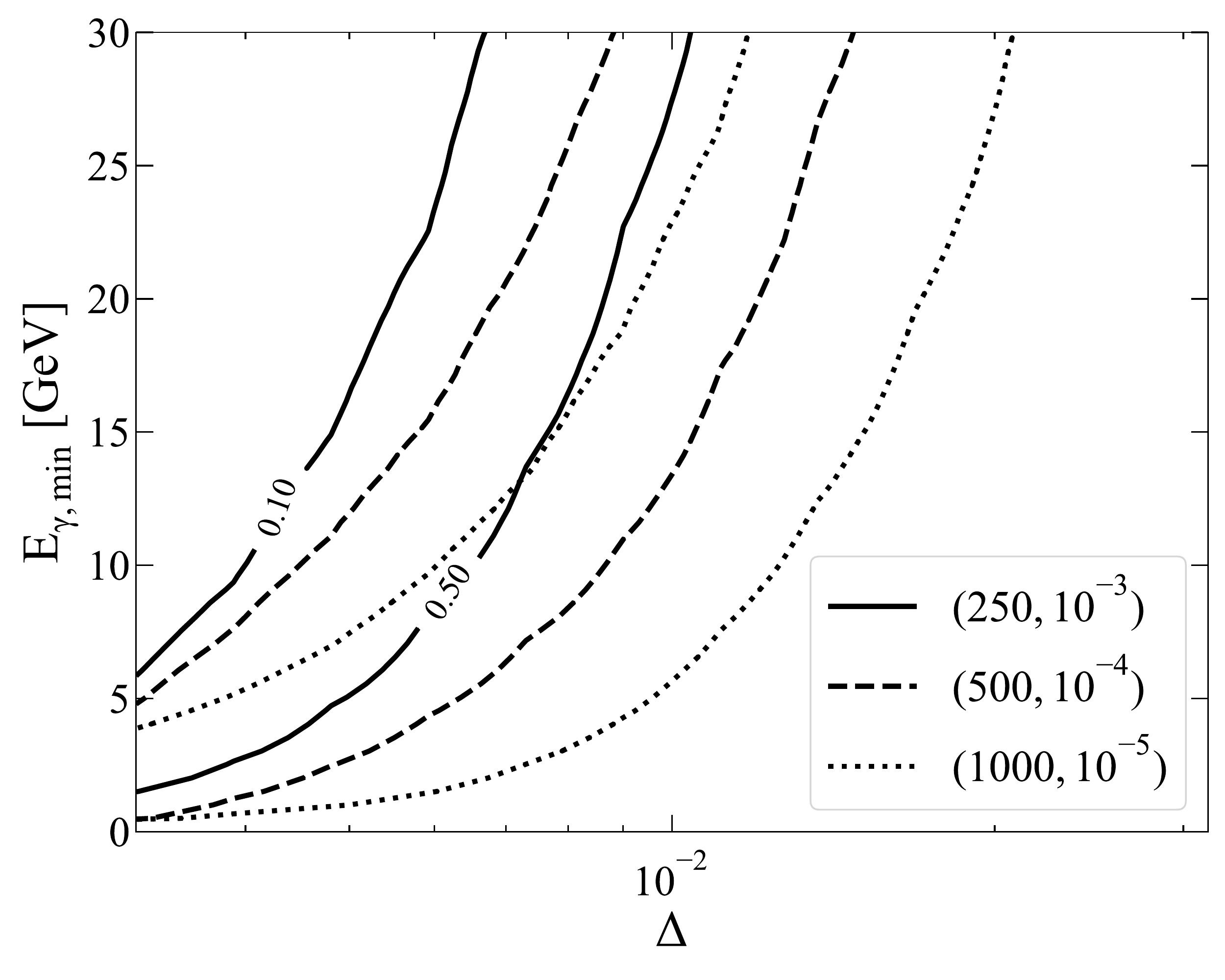}
  \includegraphics[width=0.32\textwidth]{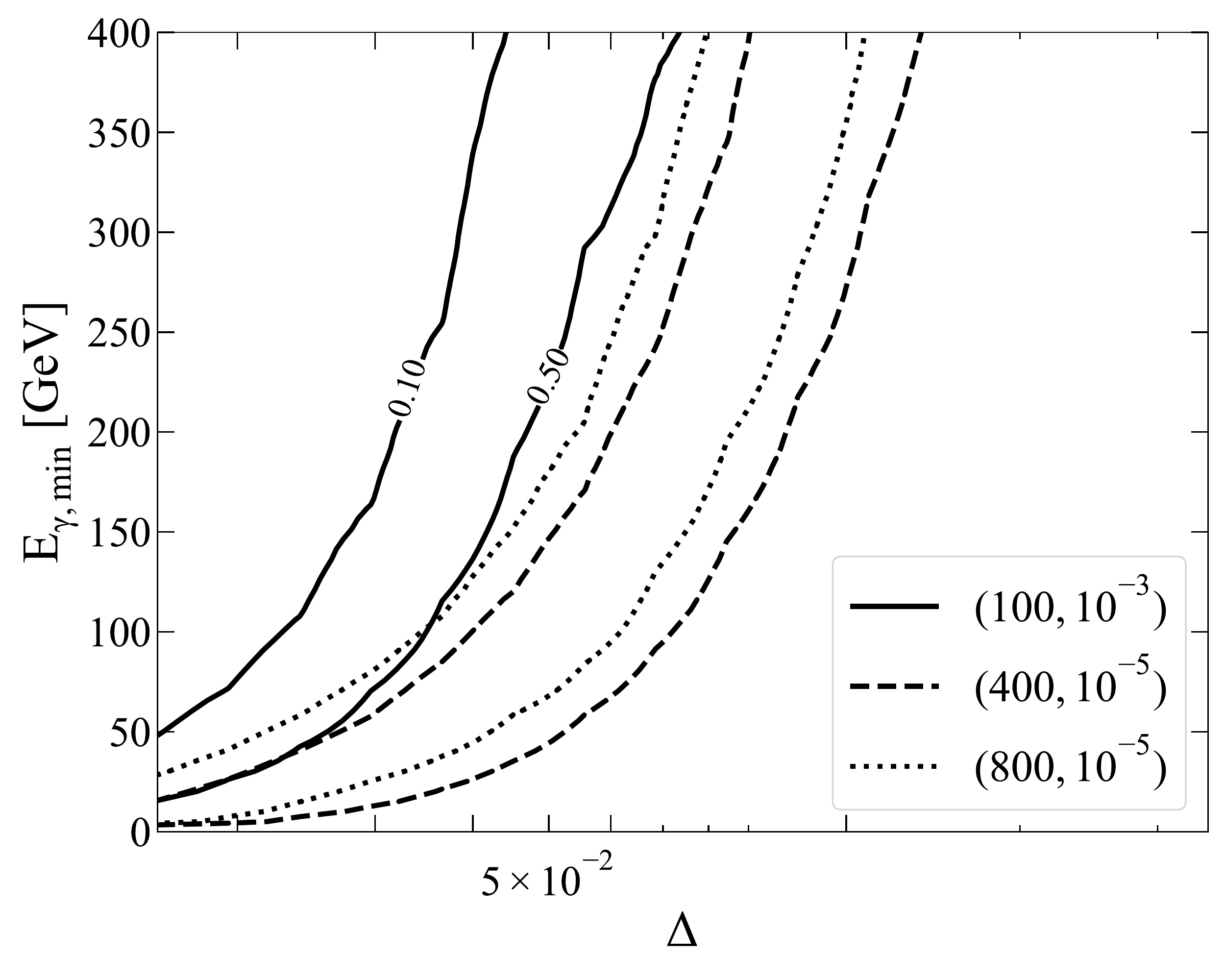} 
\caption{Contours within the $(\Delta,E_{\gamma,{\rm min}})$-plane of the fraction 
$f_\gamma(E_{\gamma,{\rm min}})$ of signal events that would be detected at 
FASER for a given photon-energy threshold $E_{\gamma,{\rm min}}$, relative to the 
number that would be detected for $E_{\gamma,{\rm min}}=0$.  The solid, dashed, 
and dotted contours appearing in each panel correspond to different 
combinations of the parameters $m_0$ and $\Lamcalo^{-1}$.  These combinations are specified  in the legend using the notation $(m_0,\Lamcalo^{-1})$, where $m_0$ and $\Lamcalo^{-1}$ are given in units of MeV and GeV$^{-1}$, respectively.  Two such contours are 
included for each parameter combination.  The upper contour corresponds to
$f_\gamma(E_{\gamma,{\rm min}}) = 0.1$, while the lower contour corresponds to 
$f_\gamma(E_{\gamma,{\rm min}})=0.5$.  The three panels of the figure simply 
display different regions of the $(\Delta,E_{\gamma,{\rm min}})$-plane in 
order to illustrate the shapes of these contours for the three $m_0$ and 
$\Lamcalo$ combinations shown therein. }
\label{fig:threshold}
\end{figure*}

Photons from $\chi_1 \rightarrow \chi_0 \gamma$ decay typically have energies 
$E_{\gamma}\sim (1~ {\rm TeV}) \Delta$ within our parameter-space region of interest.
However, the FASER and FASER2 detectors are primarily designed to look for 
$\mathcal{O}({\rm TeV})$ energy deposits, and the detection efficiency 
$\epsilon_\gamma(E_\gamma)$ for a monophoton signal with 
$E_\gamma \ll \mathcal{O}({\rm TeV})$ as a function of $E_\gamma$ is not well 
established.  In order to illustrate the impact of this efficiency on the sensitivity 
of these detectors to our signal process, we focus for simplicity on the case in which
$\epsilon_\gamma(E_\gamma)$ can be modeled in terms of a binary detection threshold 
$E_{\gamma,{\rm min}}$ --- \ie, we take
$\epsilon_\gamma(E_\gamma) = \Theta(E_\gamma - E_{\gamma,{\rm min}})$.  In 
\figref{threshold}, we display contours within the 
$(\Delta,E_{\gamma,{\rm min}})$-plane of the fraction
$f_\gamma(E_{\gamma,{\rm min}})$ of signal events that would be detected at FASER for 
given $E_{\gamma,{\rm min}}$, relative to the number that would be obtained for 
$E_{\gamma,{\rm min}}=0$.  The solid, dashed, and dotted contours appearing in each 
panel of the figure correspond to different combinations of the parameters $m_0$ and 
$\Lamcalo$.  Two such contours --- one corresponding to 
$f_\gamma(E_{\gamma,{\rm min}}) = 0.1$ and the other to 
$f_\gamma(E_{\gamma,{\rm min}}) = 0.5$ --- are included for each parameter combination.
The three panels of the figure simply display different regions of the 
$(\Delta,E_{\gamma,{\rm min}})$-plane in order to illustrate the shape of these 
contours for the three $m_0$ and $\Lamcalo$ combinations shown therein.  The
results displayed in the figure correspond to the case of a MDM interaction. However, the value of $f_\gamma(E_{\gamma,{\rm min}})$ for an MDM interaction does not differ by more than 5\% of the corresponding value obtained for an EDM interaction along the contours shown. In obtaining the results shown in Fig.~\ref{fig:resultsplots}, we have employed a cut on the photon energy of $E_{\gamma} > 300~{\rm MeV}$, which is well above expected energy deposits coming from minimum ionizing particles.


\subsection{Results}

\begin{figure*}
\centering
  \includegraphics[width=0.49\textwidth]{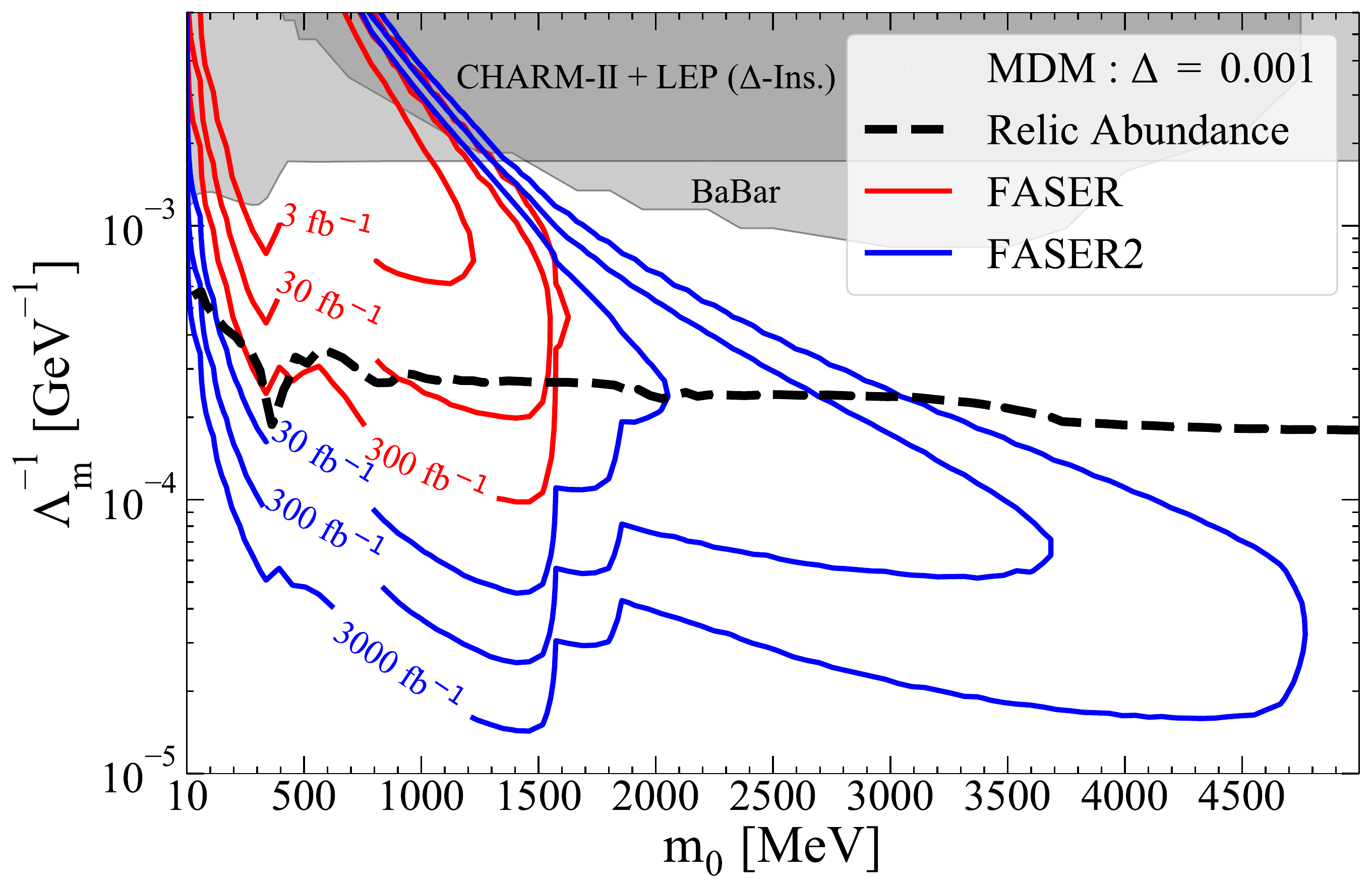}
  \includegraphics[width=0.49\textwidth]{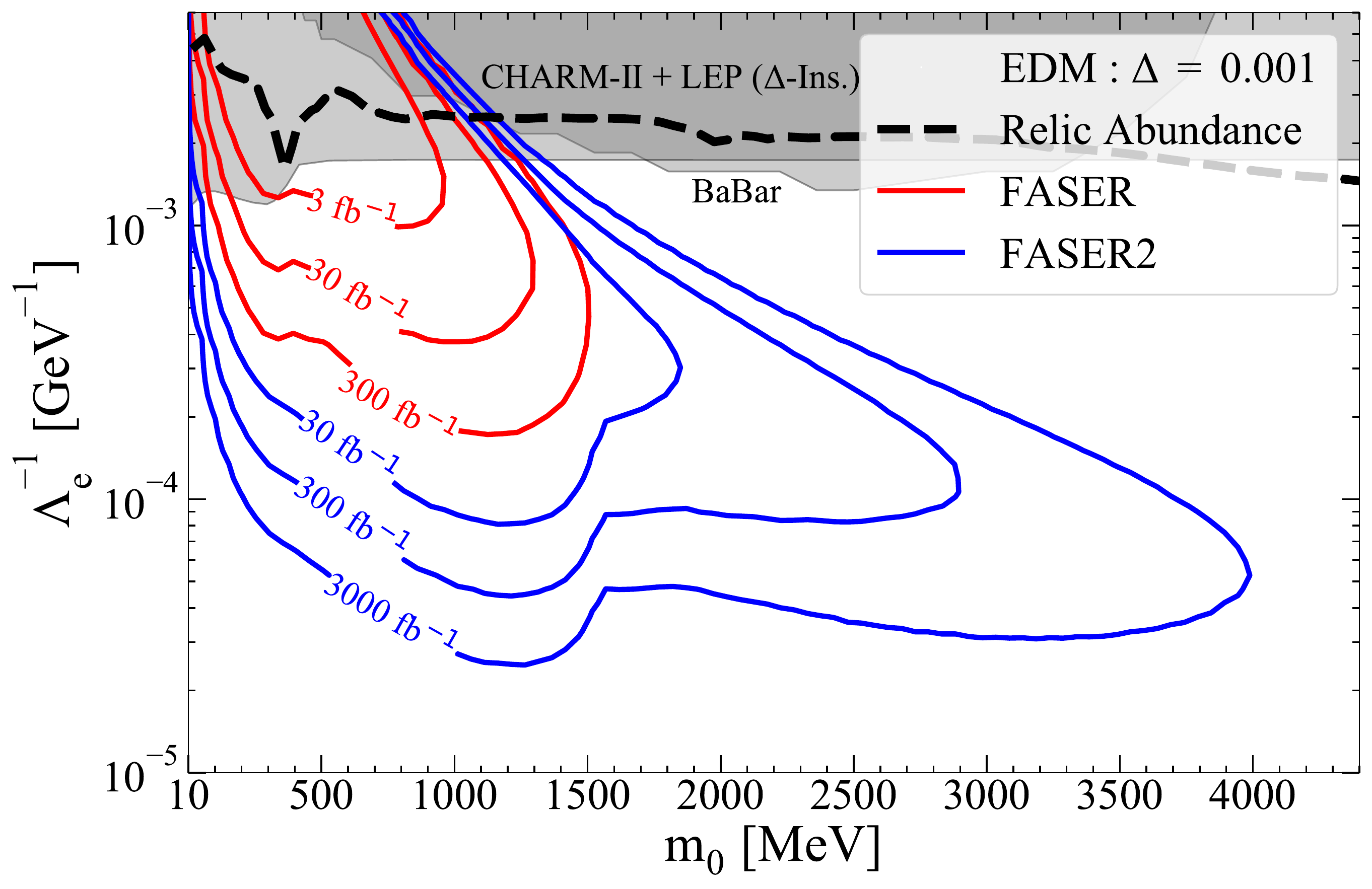}
  \includegraphics[width=0.49\textwidth]{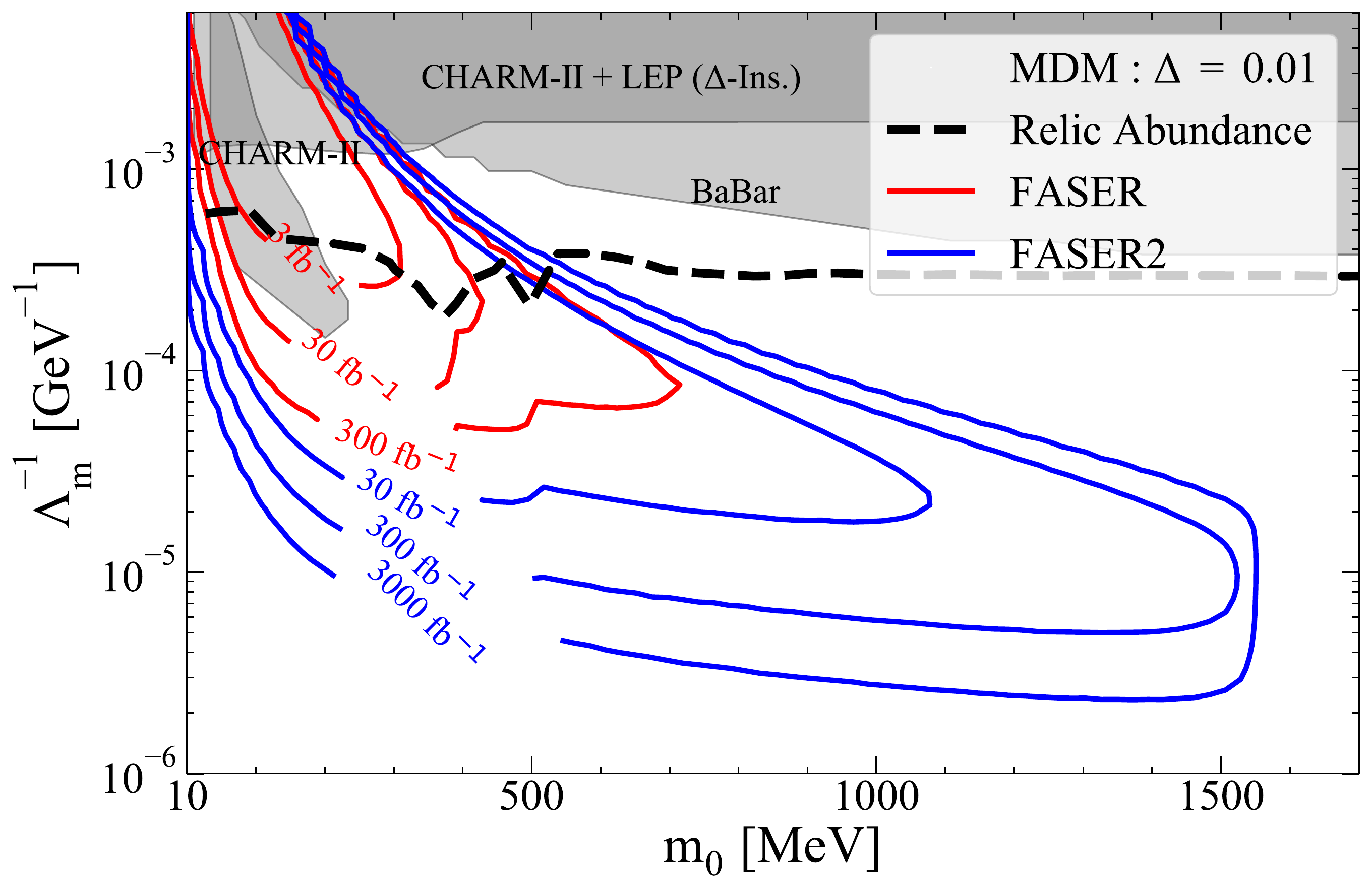}
  \includegraphics[width=0.49\textwidth]{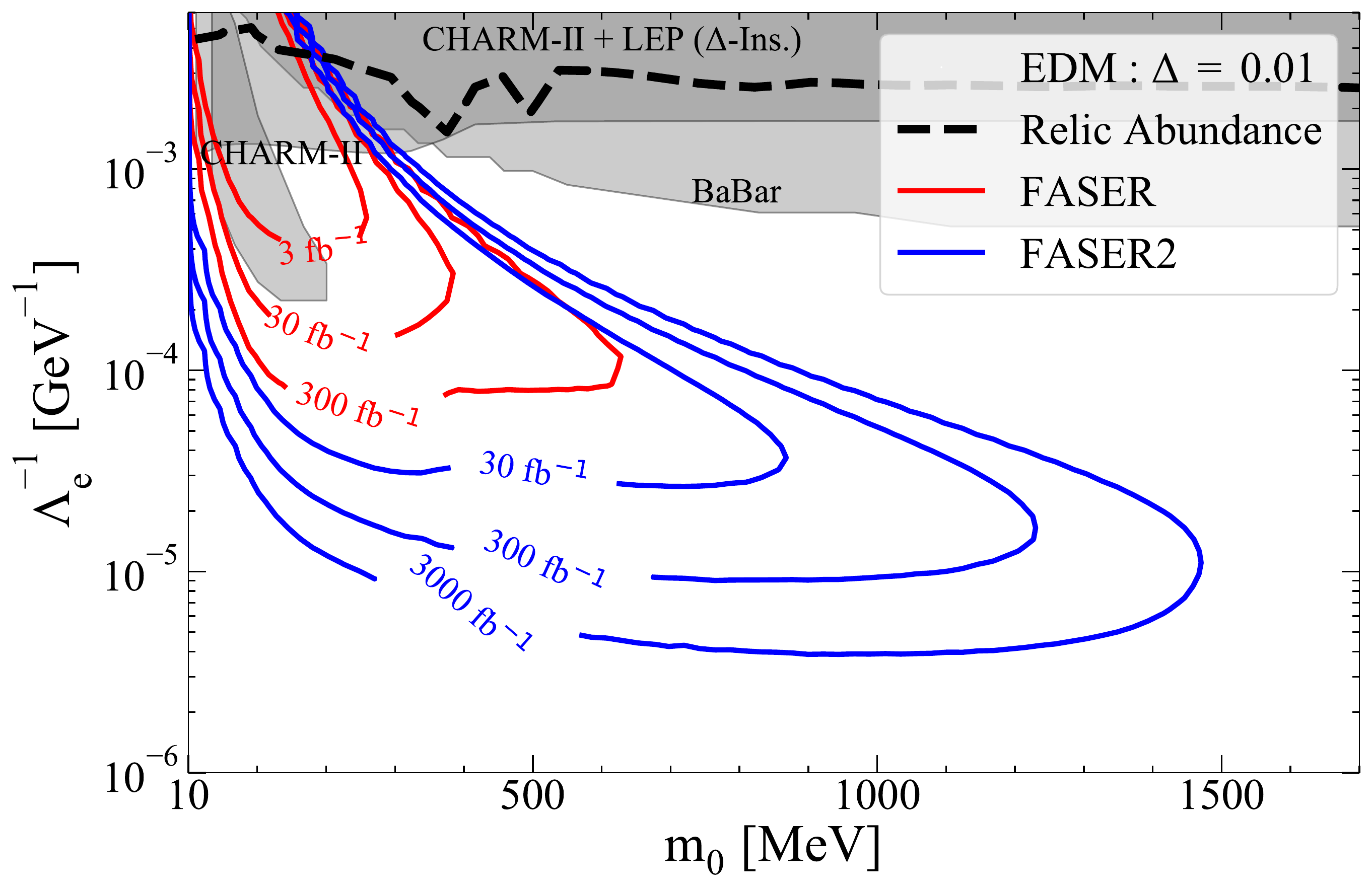}
  \includegraphics[width=0.49\textwidth]{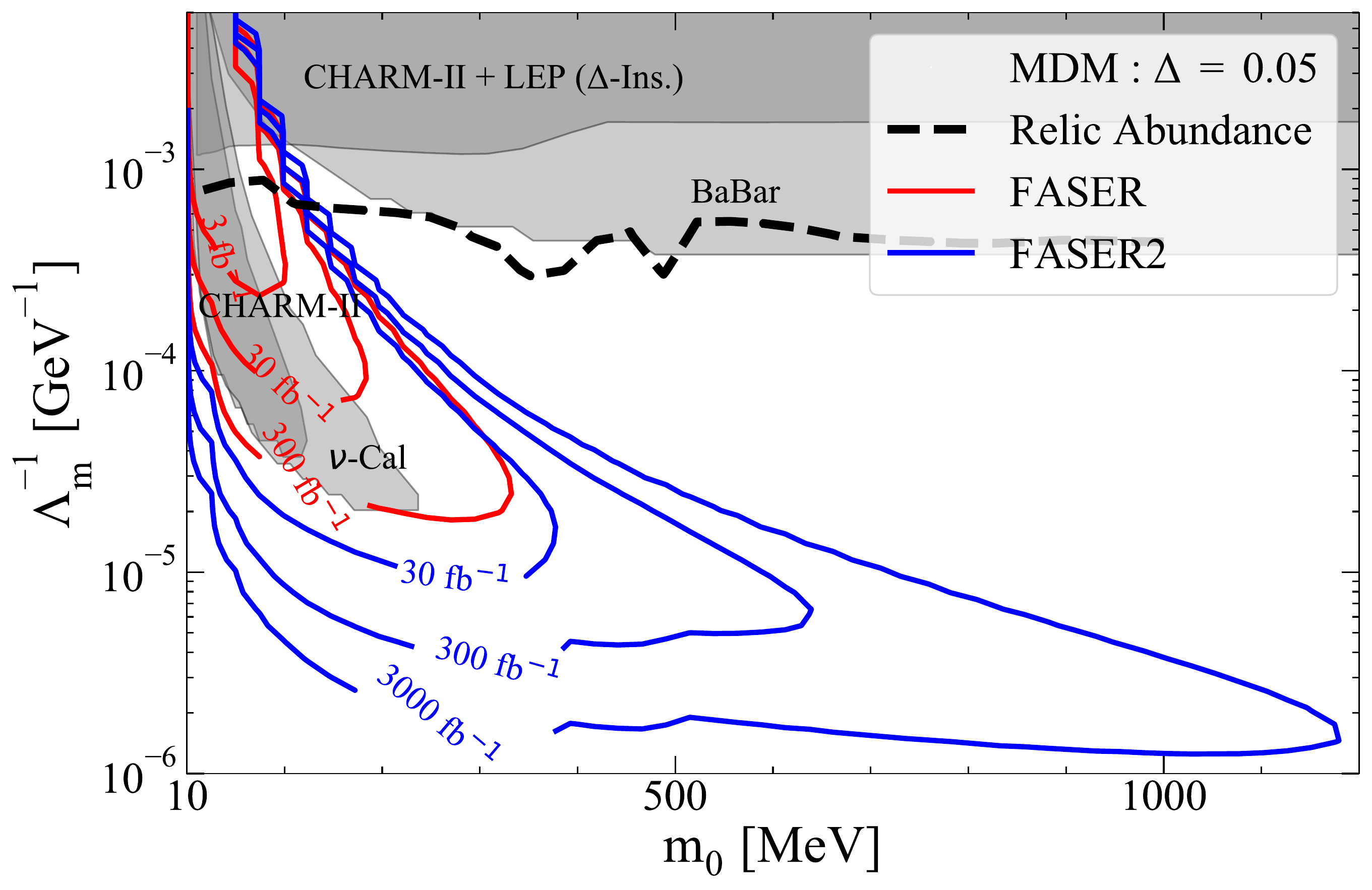}
  \includegraphics[width=0.49\textwidth]{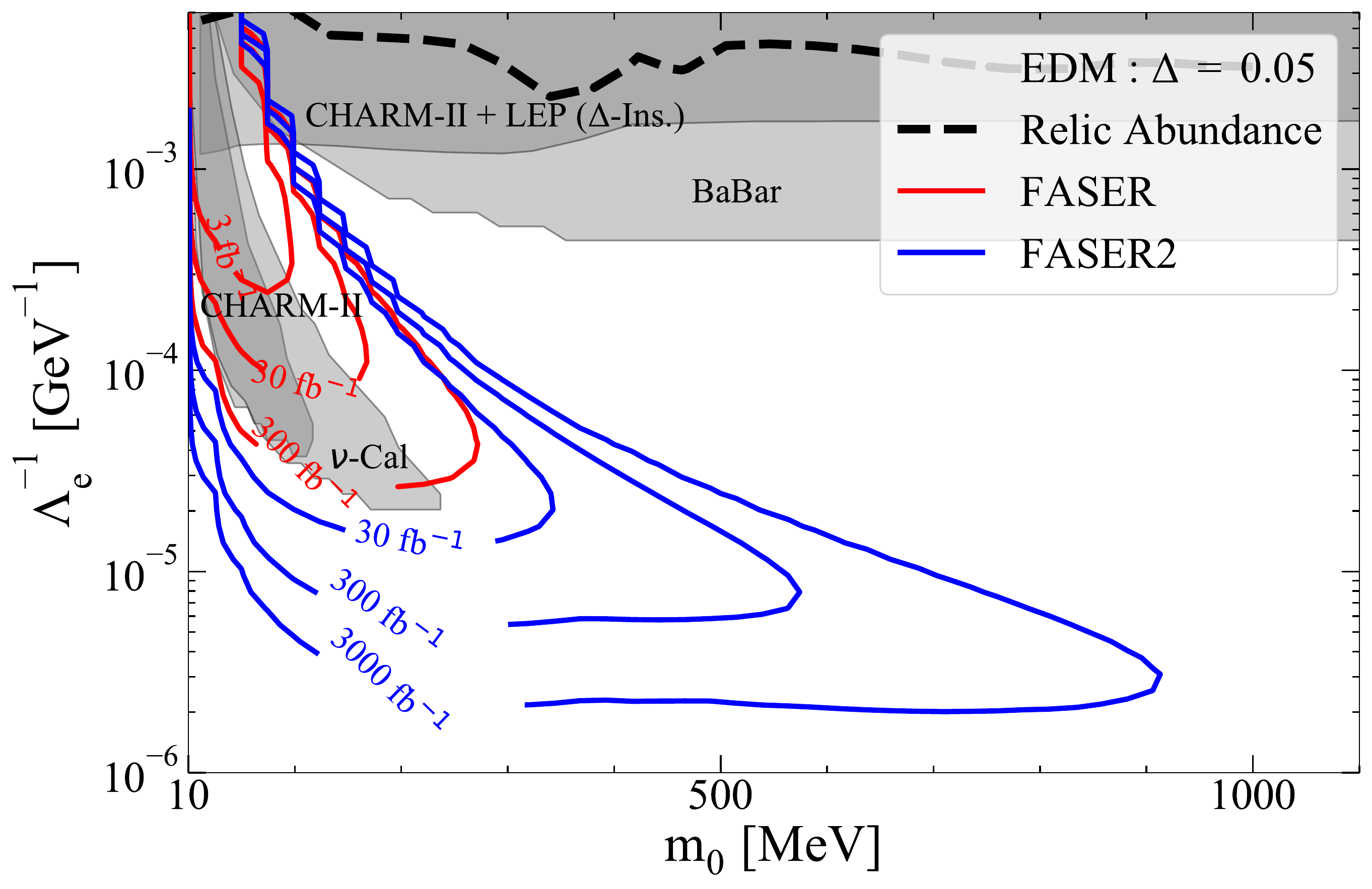}
  \caption{The integrated luminosity required to observe a monophoton signal of 
    inelastic-dipole dark matter at 95\%~C.L.\ at FASER (red curves) and FASER2 
    (blue curves) within the $(\Lamcalo^{-1},m_0)$-plane for a variety of 
    different choices of $\Delta$.  The top, middle, and bottom panels in each column 
    respectively correspond to the choices $\Delta = 0.001$, $\Delta = 0.01$, and 
    $\Delta = 0.05$, while the left and right columns respectively correspond to 
    the cases in which the dark-sector particles couple to the visible sector
    via an MDM interaction and via an EDM interaction.  The dashed curve in each panel 
    indicates the contour along which the present-day relic abundance of $\chi_0$
    accords with the observed dark-matter abundance.  The gray regions in 
    each panel are excluded by existing constraints.  These include
    constraints from searches at BaBar, constraints from Nu-Cal and CHARM-II limits 
    on $\chi_1$ decay, and constraints from probes of dipole-interacting LLPs 
    that are insensitive to the value of $\Delta$ and thus effectively identical
    to those obtained for an elastic-dipole interaction~\cite{Kling:2022ykt}.
    The exclusion region labeled ``CHARM-II + LEP ($\Delta$-Ins.)'' in each panel
    is excluded by this last set of constraints.
    \label{fig:resultsplots}}
\end{figure*}

In \figref{resultsplots}, we show contours of the integrated luminosity 
required to observe a monophoton signal of inelastic-dipole dark matter
at FASER (red curves) and FASER2 (blue curves) within the 
$(\Lamcalo^{-1},m_0)$-plane for a few of different choices of $\Delta$.
Since the expected number of background events is effectively zero, as 
discussed in \secref{background}, we adopt the observation of at least three signal 
events as our discovery criterion. We emphasize that the luminosity contours can 
also be interpreted as event contours for a fixed total luminosity, ${\cal L}_{\rm int}$ 
(cf. \tableref{Fasergeometry}), \ie~ for a contour, the number of events can be 
read as $N_{\chi_1} = 3 \times ({\cal L}_{\rm int}/{\cal L})$ , where ${\cal L}$ is 
the luminosity of the contour. This can be useful in the case where there are unexpected 
backgrounds.  The top, middle, and bottom panels in each column 
correspond to the choices $\Delta=0.001$, $\Delta =0.01$, and $\Delta = 0.05$, 
respectively. For smaller $\Delta$, we expect to be sensitive only to values of 
$\Lamcalo^{-1}$  which lie within the region excluded by constraints from CHARM-II and 
LEP --- constraints which are insensitive to $\Delta$ for $\Delta \ll 1$. While we do 
not consider smaller $\Delta$ for this reason, we note that we expect FASER also to have 
some sensitivity for smaller $\Delta$.  The results in the left column correspond to the 
case of a MDM interaction, while the results in the right column correspond to the case 
of an EDM interaction.  The dashed black curve in each panel indicates the contour along 
which the present-day relic abundance of $\chi_0$ --- a derivation of which shall 
be presented in \secref{RelicAbundance} --- accords with the dark-matter 
abundance inferred from Planck data~\cite{Planck:2018vyg}. For $\Delta \ll 1$, one 
expects our model to approach the elastic limit (\ie ~ $\Delta = 0$), and thus the relic 
target should remain fixed for $\Delta < 10^{-3}$.  The gray regions in 
each panel are excluded by existing constraints --- constraints that will 
be discussed in more detail in Sect.~\ref{sec:ExistingConstraints}.

The overall shape of the sensitivity contours that appear in \figref{resultsplots} 
can be understood as follows.  For a given integrated luminosity and a given 
combination of $m_0$ and $\Delta$, the range of $\Lamcalo$ for which a 
statistically significant number of $\chi_1$ particles decay within the FASER or 
FASER2 detector is bounded from both above and below.  The lower bound stems from
the requirement that $\bar{d}$ be sufficiently large that a non-negligible number
of $\chi_1$ particles reach the decay volume before they decay.  Taken together,
Eqs.~(\ref{eq:Shakespeare}) and~(\ref{eq:GammaChi1}) imply that
$\bar{d}\propto \Lamcalo^2/(m_0^4\Delta^3)$ for fixed $|\vec{\mathbf{p}}_{\chi_1}|$.
This implies that for a fixed value of $\Delta$, we expect the contour in the 
$(\Lamcalo^{-1},m_0)$-plane above which a significant number of $\chi_1$ particles
reach the FASER decay volume before they decay to take the form
$\Lamcalo^{-1} \propto m_0^{-2}$.  Indeed, the shape 
of the upper part of each sensitivity contour shown in each panel of
\figref{resultsplots} takes this form. 

By contrast, the upper bound on the corresponding range of $\Lamcalo$ values
for which FASER or FASER2 is sensitive for a given integrated luminosity and
a given combination of $m_0$ and $\Delta$ stems from the fact that for 
very large $\Lamcalo$, most $\chi_1$ particles that enter the decay volume
of the detector pass through it without decaying.
For $\bar{d} \gg L$, the expression for the 
decay probability in Eq.~(\ref{eq:PL}) reduces to   
$P(\bar{d},h,L)\approx h/\bar{d}$.  Moreover, the production rate of 
$\chi_1$, which is dominated by the contribution from vector-meson decay, 
is also suppressed for large $\Lamcalo$ by a factor of 
${\rm{BR}}(V\rightarrow \chi_1 \overline{\chi}_0) \propto \Lamcalo^{-2}$.
As a result, the expected number of $\chi_1$ decays within the FASER or
FASER2 detector scales with $\Lamcalo$, $m_0$, and $\Delta$ like
\begin{equation}
  N_{\chi_1} ~\propto~ {\rm{BR}}(V\rightarrow \chi_1 \overline{\chi}_0)\,
  P(\bar{d},h,L) ~\sim~ \frac{m_0^4 \Delta^3}{\Lamcalo^4}~. 
\end{equation}
Thus, for a fixed value of $\Delta$, the maximum value of $\Lamcalo$ for 
which $N_{\chi_1}$ drops below the sensitivity threshold scales with $m_0$
according to a proportionality relation of the form $\Lamcalo \propto m_0$.
This scaling behavior accounts for the shape of the lower part of each 
sensitivity contour shown in the figure. 

The ``bumps'' that appear in these sensitivity contours --- bumps which are 
the most prominent in the upper left panel --- arise as a consequence of the decay 
kinematics of the individual meson species that contribute to the production 
of $\chi_1$.  In particular, as $m_0$ increases, mesons with masses 
$m_M < (2+\Delta)m_0$ become kinematically forbidden from decaying into
$\chi_i$ pairs, and as a result the overall production rate of $\chi_1$ 
particles decreases.

Comparing the results shown in the corresponding panels in the left and right 
columns of the figure, we observe that the discovery reach obtained for 
an MDM operator structure at both FASER and FASER2 is comparable to, but
nevertheless slightly larger than, the reach obtained for an EDM operator
structure.  This is a reflection of the fact that as $m_0$ approaches the kinematic
cutoff $m_0 = m_V/(2+\Delta)$ above which the decay process 
$V\rightarrow \chi_1\chi_0$ is kinematically forbidden for a given 
vector-meson species, $\sigma_{\chi_1 M}^{({\rm eff})}$ falls off more 
rapidly for an EDM operator structure than it does for a MDM operator
structure, as indicated in \figref{spectra}.

The primary message of \figref{resultsplots}, however, is that both FASER
and FASER2 are capable of probing a sizable region of the parameter space
of inelastic-dipole-dark-matter models that has not been meaningfully probed by 
other experiments.  As one would expect, the discovery reach afforded by FASER2 
within that parameter space is significantly larger than that  
afforded by FASER.~  However, we observe that both experiments are
sensitive within hitherto unprobed regions of that parameter space wherein
the $\chi_0$ particle has the correct relic abundance to account for the
entirety of the dark matter.  Moreover, we emphasize that while regions
of parameter space that lie below the black dashed line would overproduce
dark matter in the context of the standard cosmology, such regions of
parameter space would be viable in modified cosmological scenarios involving,
for example, additional entropy production from the out-of-equilibrium decays 
of heavy particles after thermal freeze-out, but before the nucleosynthesis 
epoch.


\section{Relic Abundance\label{sec:RelicAbundance}}


In order for the $\chi_i$ to play the role of the dark matter in our 
inelastic-dipole scenario, the collective present-day abundance of these states 
must be consistent with the dark-matter abundance 
$\Omega_{\rm DM} \approx 0.26$ inferred from Planck data~\cite{Planck:2018vyg}.
Thermal freeze-out provides a natural mechanism for generating abundances 
for both $\chi_0$ and $\chi_1$ in this scenario.  
Within the region of parameter space relevant for FASER, $\chi_1$ is sufficiently 
short-lived that essentially all $\chi_1$ particles present in the universe at 
the end of the freeze-out epoch decay to $\chi_0$ particles well before 
present time --- and for that matter, well before the Big-Bang Nucleosynthesis (BBN)
epoch.  Thus, within this region of parameter space, only $\chi_0$ contributes 
to the present-day abundance of dark matter.  Nevertheless, as we shall see, 
co-annihilation processes involving $\chi_1$ particles during the freeze-out epoch 
can have a significant impact on the resulting $\chi_0$ abundance at late times.

The evolution of the number densities $n_i$ of the $\chi_i$ in the early universe
is governed by set of coupled Boltzmann equations, each of which takes the form
\beq
  \frac{dn_i}{dt}+3Hn_i ~=~ C_i~,
  \label{eq:Boltzmann_eqs}
\eeq
where $H$ is the Hubble parameter and where $C_i$ represents the contribution to the 
rate of change of $n_i$ from scattering and decay processes.  Equivalently,
these differential equations may be formulated in terms of the 
comoving number densities $Y_i \equiv n_i/s$ of the dark-sector species, where 
$s$ denotes the entropy density of the universe.  Indeed, during periods wherein the 
total entropy $S \propto sa^3$ within a comoving volume is conserved, 
the Boltzmann equations for the $Y_i$ each take the particularly simple form 
\beq
  \frac{dY_i}{dt} ~=~ \frac{C_i}{s}~.
  \label{eq:yield}
\eeq

In inelastic-dipole-dark-matter models, the processes that can contribute 
significantly to the $C_i$ include the following, where $\ell = e,\mu,\tau$
denotes a charged-lepton species:
\begin{itemize}
\setlength\itemsep{-0.2em}
 \setlength{\itemindent}{-1.5em}
 \item $t$-channel annihilation processes 
      $\overline{\chi}_i \chi_i\leftrightarrow \gamma \gamma$;
    \item $s$-channel co-annihilation processes 
      $\overline{\chi}_i \chi_j \leftrightarrow \ell^+ \ell^-$ and 
      $\overline{\chi}_i \chi_j\leftrightarrow \mbox{hadrons}$,
      where $i\neq j$;
    \item up- and down-scattering processes 
      $\chi_0 \ell^\pm\leftrightarrow \chi_1 \ell^\pm$ and
      analogous processes involving hadrons; 
    \item decay and inverse-decay processes 
      $\chi_1\leftrightarrow \chi_0 \gamma$.
\end{itemize}
The annihilation and co-annihilation processes listed above can 
change the {\it total}\/ comoving number density $Y_{\rm tot} \equiv Y_0 + Y_1$ of 
dark-sector particles.  By contrast, the other processes listed above can transfer 
comoving number density from one dark-sector species to the other, but have no effect 
on $Y_{\rm tot}$.   

We can express the  $C_i$ for each $\chi_i$ as a sum of four terms, each associated 
with one of the four classes of process itemized above.  In particular, $C_0$ and $C_1$ 
may be written in the form
\begin{eqnarray}
  C_0 &~=~& 
      -\left[ {n_0}^2 \! - \! \left({n_0^{\rm eq}}\right)^2\right]
      \expt{\sigma_{00}v} - \Big(n_0n_1-n_0^{\rm eq}n_1^{\rm eq}\Big)
      \expt{\sigma_{01}v}\nn\\
    & & - \! \left( \! n_0 \! - \! \frac{n_1}{n_1^{\rm eq}}n_0^{\rm eq} \! \right)
      n_{e}^{\rm eq}\expt{\sigma_{0e}v}
       \! + \! \left( \! n_1 \! - \! \frac{n_0}{n_0^{\rm eq}}n_1^{\rm eq} \! \right) 
       \! \expt{\Gamma} 
       \nonumber \\
  C_1 &~=~& 
      -\left[ {n_1}^2 \! - \! \left({n_1^{\rm eq}}\right)^2\right]\expt{\sigma_{11}v}
      -\Big(n_0n_1-n_0^{\rm eq}n_1^{\rm eq}\Big)\expt{\sigma_{01}v} \nn \\
    & & + \! \left( \! n_0 \! - \! \frac{n_1}{n_1^{\rm eq}}
    n_0^{\rm eq} \! \right)n_{e}^{\rm eq}\expt{\sigma_{0e}v}
     \! - \! \left( \! n_1 \! - \! \frac{n_0}{n_0^{\rm eq}}n_1^{\rm eq} \! \right) 
     \! \expt{\Gamma}~,
     \nonumber \\
  \label{eq:collision}
\end{eqnarray}
where $n_i^{\rm eq}$ is the equilibrium number density of $\chi_i$, where
$n_\ell^{\rm eq}$ is the equilibrium number density of the charged-lepton species 
$\ell$, and where $\expt{\Gamma}$ denotes the thermally-averaged decay rate for $\chi_1$. 
In addition, $\expt{\sigma_{00}v}$, $\expt{\sigma_{11}v}$, $\expt{\sigma_{12}v}$, 
and $\expt{\sigma_{0\ell}v}$ denote the thermally-averaged cross-sections 
for the annihilation of two $\chi_0$ particles, the annihilation of two $\chi_1$ 
particles, the co-annihilation of a $\chi_0$ and a $\chi_1$ particle, and the 
up-scattering process $\chi_0 \ell^\pm \to \chi_1 \ell^\pm$, respectively.
Expressions for the thermally-averaged cross-sections and decay widths for all
relevant processes are provided in Appendix~\ref{app:XSecs}.

Within our parameter-space region of interest, $\chi_0$ and 
$\chi_1$ always depart from thermal equilibrium only after they have both 
become non-relativistic.  Within this region of interest, whether or not 
co-annihilation processes have a significant impact on the overall dark-matter
abundance depends primarily on the value of $\Delta$.  When $\Delta \gtrsim 1$, 
co-annihilation processes are relatively unimportant because the number density of 
$\chi_1$ will be exponentially suppressed relative to that of $\chi_0$ at the 
time when $\chi_1$ effectively becomes non-relativistic.  As a result, 
the cosmological evolution of $\chi_1$ is essentially decoupled from that of 
$\chi_0$.  By contrast, when $\Delta \ll 1$, the number 
densities of $\chi_0$ and $\chi_1$ remain comparable until much later, and the 
effect of co-annihilation processes cannot be ignored.  Indeed, within the
region of parameter space relevant for FASER, wherein $\Delta \ll 1$, 
co-annihilation processes turn out to play a significant role --- and indeed often
a dominant role --- in determining the overall present-day abundance of $\chi_0$. 

Of course, the freeze-out dynamics in this scenario, and therefore the late-time 
abundance of $\chi_0$, depends not only on $\Delta m$, but on $m_0$ as well.  
Since the temperature at which thermal freeze-out occurs is typically 
$m_0/\mathcal{O}(10)$, for small masses $m_i\ll m_\mu$ the production of 
$\chi_i$ involves only electrons and positrons.  However, for $m_i\gtrsim m_\mu$,
processes involving muons are also relevant.  At even higher temperatures, 
interactions involving hadrons or tau leptons become relevant as well.  

At first glance, it might seem that all of the collision terms in 
Eq.~(\ref{eq:collision}) contribute to maintaining thermal equilibrium 
between the $\chi_i$ and the thermal bath of SM particles, given that all 
of these terms vanish when $n_i=n_i^{\rm eq}$.  Were this the case, the 
up- and down-scattering process in which the $\chi_i$ participate, as well 
as the decay and inverse-decay processes, would serve to keep $Y_0$ and $Y_1$ 
in equilibrium for a significant duration, given that the 
number densities of the light SM particles that participate in these 
processes are still quite large during the freeze-out epoch.  As a result,  
the relic abundance of $\chi_0$ would be heavily suppressed.
However, as we have discussed above, these processes --- unlike annihilation 
and co-annihilation processes --- have no effect on $Y_{\rm tot}$.  Thus,
after the annihilation and co-annihilation rates drop below $H$, 
both $\chi_0$ and $\chi_1$ depart from the thermal equilibrium with 
the SM thermal bath.  Nevertheless, the up- and down-scattering processes, 
together with the decay and inverse-decay processes, serve to maintain
the relations
\begin{eqnarray}
  \frac{Y_0}{Y_1} &~=~& \frac{Y_0^{\rm eq}}{Y_1^{\rm eq}}
  \label{eq:ratio_condition} \nonumber \\
  Y_{\rm tot}&~\equiv~&Y_0+Y_1~\approx ~ {\rm const.}
  \label{eq:}
\end{eqnarray}
between $Y_0$ and $Y_1$ even after this departure from equilibrium occurs.
These relations imply that
\beqn
  Y_0 &~=~& \frac{Y_0^{\rm eq}}{Y_0^{\rm eq}+Y_1^{\rm eq}}\,Y_{\rm tot}
    \nonumber\\
  Y_1 &~=~& \frac{Y_1^{\rm eq}}{Y_1^{\rm eq}+Y_0^{\rm eq}}\,Y_{\rm tot}~.
\eeqn
Thus, at late time, when $T\ll m_1-m_0$ and the equilibrium abundances 
therefore satisfy $Y_1^{\rm eq}\ll Y_0^{\rm eq}$, it therefore follows
that
\begin{eqnarray}
  Y_0 &~\approx~& Y_{\rm tot} \nonumber\\ 
  Y_1 &~\approx~& \left(\frac{m_1}{m_0}\right)^{3/2}
    \exp\left(-\frac{m_1-m_0}{T}\right)Y_{\rm tot}~.
  \label{eq:Y0Y1LowT}
\end{eqnarray}

Physically, the results in Eq.~(\ref{eq:Y0Y1LowT}) reflect the fact that
even if $Y_0$ and $Y_1$ are comparable at the moment when 
$\chi_0$ and $\chi_1$ initially depart from thermal equilibrium with the radiation bath,
up-scattering and inverse-decay processes will become increasingly kinematically 
disfavored as $T$ drops.  As a result, the initial population of $\chi_1$ particles
present at the point at which this departure occurs will gradually be converted into 
$\chi_0$ particles via decay and down-scattering processes until the majority 
of this initial population is all but depleted and $\chi_0$ freezes out.  

\begin{figure}
    \centering
    \includegraphics[width=0.48\textwidth]{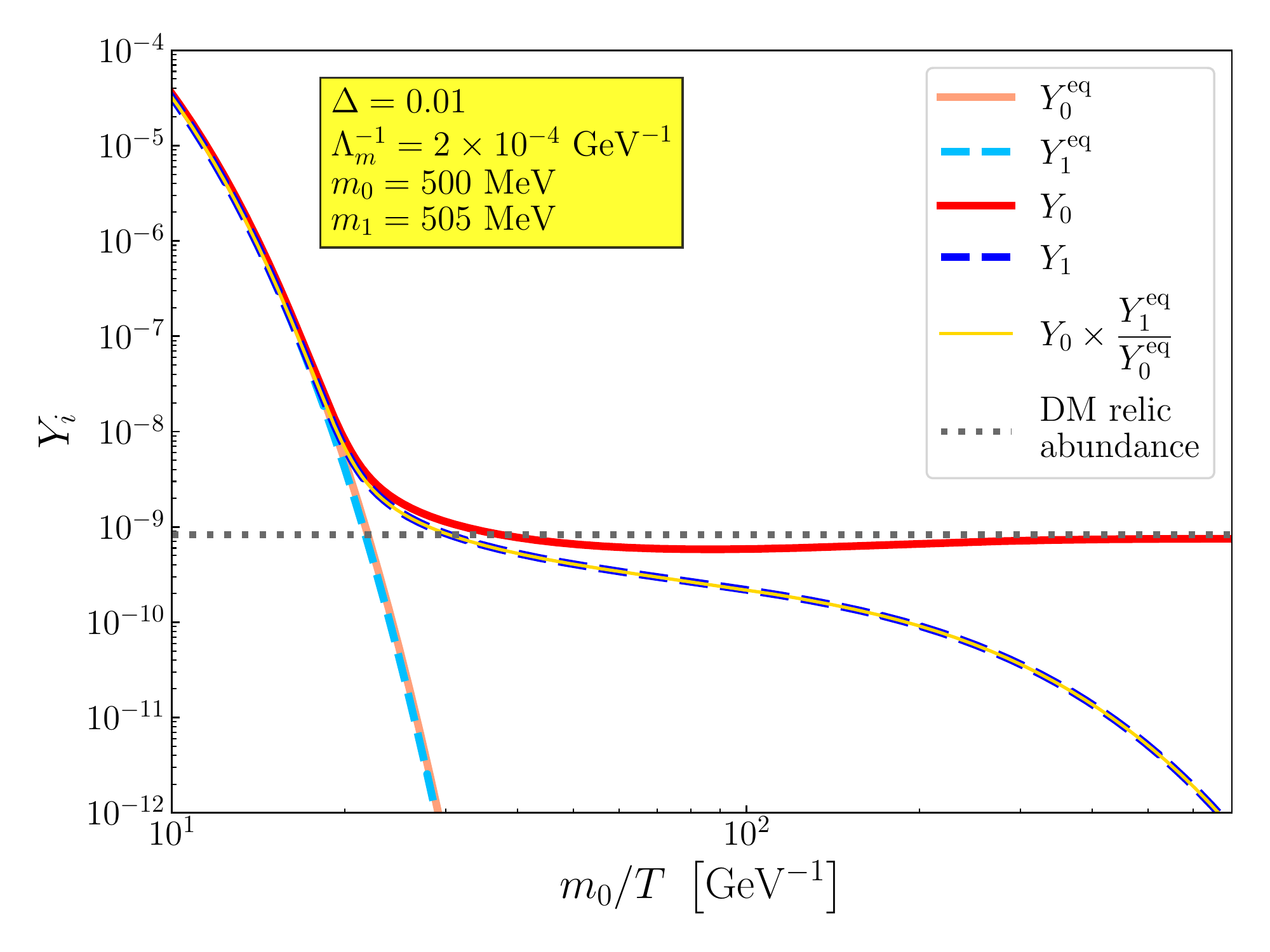}
    \caption{The evolution of the comoving number densities $Y_i$ of the two 
      dark-sector fields $\chi_i$, shown as functions of $m_0/T$ for the parameter 
      assignments specified in the legend.
      The solid red and dashed blue curves respectively represent $Y_0$ and 
      $Y_1$, whereas the solid orange and dashed cyan curves represent the corresponding 
      equilibrium comoving number densities $Y_0^{\rm eq}$ and $Y_1^{\rm eq}$. The 
      dotted gray line indicates the value of $Y_i$ that yields the correct
      present-day dark-matter abundance for this choice of model parameters.
      The yellow curve represents the quantity $Y_0 (Y_1^{\rm eq}/Y_0^{\rm eq})$,
      which, in accord with Eq.~(\ref{eq:ratio_condition}), 
      effectively coincides with $Y_1$ across the entire range of $m_0/T$ shown.}
    \label{fig:yield}
\end{figure}

In Fig.~\ref{fig:yield}, we show how the comoving number densities $Y_0$ and 
$Y_1$ evolve as a function of $m_0/T$ for an example choice of model 
parameters that yields a relic abundance for $\chi_0$ that accords with 
$\Omega_{\rm DM}$.  Since $\Delta \ll 1$ for this choice of parameters,
the equilibrium comoving number densities $Y_0^{\rm eq}$ and $Y_1^{\rm eq}$ 
(indicated by the solid orange and dashed cyan curves in the figure) are very 
similar.  As a result, the actual comoving number densities $Y_0$ and $Y_1$,
which track their equilibrium values very closely at high temperatures, likewise 
remain quite similar at high temperatures.  However, at around $m_0/T\lesssim 20$, 
when $Y_0$ and $Y_1$ begin to depart appreciably from their equilibrium values, 
they also begin to differ appreciably from each other.  Indeed, by the time at 
which $Y_0$ has effectively settled into its late-time asymptotic value --- a time 
that corresponds roughly to the point at which $m_0/T\sim 200$ --- this comoving 
number density is already an order of magnitude larger than $Y_1$.  We 
also observe that $Y_1$ tracks the quantity $Y_0 (Y_1^{\rm eq}/Y_0^{\rm eq})$, 
in accord with the result in Eq.~(\ref{eq:ratio_condition}).

We also note that the value of $\Lambda_{\mathcal{O}}$ required to obtain the 
correct thermal relic density in the EDM case is smaller than it is in 
the MDM case for the same choice of $m_0$ and $\Delta$.  This is ultimately 
a consequence of angular-momentum conservation.  In the regime in which
$\Delta \ll 1$ and $\chi_0$ and $\chi_1$ are nearly degenerate 
in mass, the thermal relic density is determined primarily by co-annihilation 
processes such as $\overline{\chi}_0 \chi_1 \to e^+ e^-$, which are mediated 
by an $s$-channel photon.  In the MDM case, these are $S$-wave processes: 
as noted in Sect.~\ref{sec:chiproduction}, the initial state has spin $S=1$, 
and since the final state also has spin $S=1$, angular-momentum 
conservation can be satisfied by states with orbital angular momentum 
$L=0$.  By contrast, in the EDM case, they are $P$-wave processes:
the initial state has spin $S=0$ and the final state has spin $S=1$, so 
angular-momentum conservation requires orbital angular momentum $L=1$.
The value of $\Lambda_{\mathcal{O}}$ must therefore be smaller in the EDM case
to compensate for the resulting $P$-wave suppression. 

We can obtain a more quantitative picture of how this $P$-wave suppression
affects the freeze-out dynamics in the EDM case relative to the MDM case by 
comparing the corresponding expressions for the cross-section 
$\sigma_{01 \to ee}^{\rm MDM}$ and 
$\sigma_{01 \to ee}^{\rm EDM}$ for the co-annihilation process 
$\overline{\chi}_0 \chi_1 \to e^+ e^-$ --- expressions respectively obtained by
substituting the $C^{(s)}$ factors in Eqs.~(\ref{eq:CeeSExprMDM}) 
and~(\ref{eq:CeeSExprMDM}) into Eq.~(\ref{eq:eeSExpr}).  In particular, since 
$\Delta \ll 1$ within our parameter-space region of interest, one can gain some 
insight into how the differences between these expression affect our results 
by considering how these expressions behave in the $\Delta \rightarrow 0$
limit.  In this limit, one finds that these expressions reduce to
\begin{eqnarray}
  \sigma_{01 \to ee}^{\text{MDM}} &~=~& 
    \frac{e^2}{6 \pi \Lambda_m^2} \left(\frac{3 - 2 v^2 }{v}\right) 
    \nonumber\\
  \sigma_{01 \to ee}^{\text{EDM}} &~=~& 
    \frac{e^2}{6 \pi \Lambda_e^2} \, v \ ,
\end{eqnarray}
where $v = \sqrt{1 - 4m_0^2/s}$ is the speed of either initial-state particle in 
the center-of-mass frame.  Thus, we see that the co-annihilation cross-section in the 
EDM case is suppressed relative to the co-annihilation cross-section in the 
MDM case by a factor of $v^2$.  At freeze-out, when $v^2 \sim 0.1$, the two 
cross-sections are equal when $\Lambda_m \sim 5 \Lambda_e$.  Indeed, this is 
reflected in the locations of the $\Omega_{\chi_0} = \Omega_{\rm DM}$ contours 
appearing in \figref{resultsplots}.


\section{Existing Constraints\label{sec:ExistingConstraints}}


The parameter space of inelastic-dipole-dark-matter models is constrained
by a variety of experimental and observational considerations. 
Within the region of that parameter space that we have identified as being the 
most relevant for FASER, the most important such constraints are those from 
colliders and from fixed-target experiments.  In addition, in cases in which 
the present-day abundance of $\chi_0$ particles is non-negligible, astrophysical 
considerations also imply constraints on that region of parameter space.  We 
discuss these constraints in turn below.

\subsection{Decay Signatures:~Colliders\label{sec:colliders}}

The BaBar experiment at the PEP-II $e^+ e^-$ facility is sensitive to regions of 
parameter space near the thermal-relic contour for inelastic-dipole dark matter 
and indeed can be seen as complementary to FASER.~  The resulting constraints on this 
parameter space were investigated in Ref.~\cite{Izaguirre:2015zva} for the case of 
a MDM interaction.  We extend this analysis here both to the case of an
EDM interaction and to smaller values of $\Delta$.  

Limits on events involving either one or two energetic photons and sizable missing 
energy $\slashed{E}$ at BaBar place constraints on inelastic-dipole dark matter.
The dark-sector particles $\chi_0$ and $\chi_1$ can be produced there either through the 
$t$-channel process $e^+ e^-\rightarrow \gamma \chi_1 \overline{\chi}_0$ 
or through the $s$-channel process $e^+ e^-\rightarrow \chi_1 \overline{\chi}_0$.
The leading contribution to the signal-event rate in the 2$\gamma + \slashed{E}$ channel 
arises from $t$-channel production, followed by the prompt 
decay of the $\chi_1$.  By contrast, sizable contributions to the signal-event rate in the 
$\gamma + \slashed{E}$ channel arise both from $s$-channel production, followed by the 
prompt decay of the $\chi_1$, and from $t$-channel production, followed by the  
$\chi_1$ escaping the detector before it decays. 

For our analysis of the signal rate in both the $2\gamma + \slashed{E}$ and  
$\gamma + \slashed{E}$ channels, we make use of the BaBar monophoton trigger,
which was in effect while 60~fb$^{-1}$ of integrated luminosity was accumulated.
This trigger requires that the leading photon in the event have energy 
$E_\gamma\geq2$~GeV.~  For both channels, we also require that 
$\slashed{E} \geq 20$~MeV.~  For the $2\gamma + \slashed{E}$ channel,
following Refs.~\cite{Izaguirre:2015zva,Chu:2018qrm}, we also require that an 
additional photon with $E_\gamma \geq 50$~MeV be present in the event.

For the $2\gamma + \slashed{E}$ channel, we assume no background and 
require that the $\chi_1$ decays promptly (\ie, within 1~cm of the 
interaction point). Since any $\chi_1$ particles which might have been detected at BaBar would necessarily have  been produced with much smaller boosts, this prompt decay length is the one relevant for our parameter-space region of interest.  By contrast, sizable SM backgrounds to the 
$\gamma + \slashed{E}$ signature arise from events involving 
neutrino production via neutral-current interactions and from 
events in which a photon is produced in conjunction with additional 
photons or $e^+e^-$ pairs that escape detection.   We find that the 
constraints derived from the $\gamma + \slashed{E}$ channel turn out to be 
subdominant to the constraints from the $2\gamma +\slashed{E}$ channel, 
and thus we focus on these constraints in what follows.

The signal contribution in the $2\gamma +\slashed{E}$ channel involves 
the prompt decay of the $\chi_1$, and so the corresponding constraints on 
our model-parameter space extend to arbitrarily small $\Lamcalo$.  These 
constraints exclude the shaded region of model-parameter space 
labeled ``BaBar'' in each panel of \figref{resultsplots}. 
We note that the Belle-II experiment~\cite{Belle-II:2010dht} at SuperKEKB,
which is currently taking data, will be able to extend the reach of 
B-factory experiments within this parameter space to larger 
$\Lamcalo$ in the near future.

Since $\chi_1$ can also be produced via analogous processes at hadron
colliders, LHC searches likewise place constraints on the parameter 
space of inelastic-dipole-dark-matter models.  The extent to which such
searches are capable of probing the parameter space of such models was 
investigated in Ref.~\cite{Izaguirre:2015zva} for the case of a MDM 
interaction.  While the results of this analysis indicate that LHC searches 
are in general capable of probing that parameter space down to
$\Lamcalo \sim \mathcal{O}(10^3)$~GeV for a broad range of $m_0$ and $\Delta$, 
they also indicate that for $\Delta \lesssim 0.05$, the region of parameter 
space that these searches are capable of probing does not extend beyond 
the region that is already excluded by BaBar data.  Thus, the constraints from
LHC searches on the region of that parameter space relevant for FASER 
are subdominant to those from BaBar and need not be considered further. 

One could also consider constraints from the LEP experiment constraining invisible 
$Z$ decays. Such constraints were considered in Ref.~\cite{Izaguirre:2015zva}, where 
the authors explored a similar inelastic dipole operator, but with a coupling to the 
hypercharge gauge boson instead of the photon. For the magnetic dipole case, they derived a 
lower bound $\Lambda_m \gtrsim 10^{3}~{\rm GeV}$ on the effective scale associated 
with the corresponding operator. Depending on the UV theory, a similar constraint might 
be applicable to our model as well. However since the $\chi_i$ have no direct coupling 
to the physical $Z$ boson in the effective theory we consider in this paper, we do not 
consider this constraint further. 

\subsection{Decay Signatures: Fixed-Target Experiments\label{sec:fixed target}}

A variety of experiments in which a beam of SM particles --- typically 
electrons or protons --- collides with a fixed target located some distance
in front of a particle detector impose non-trivial constraints on 
inelastic-dipole-dark-matter models.  Indeed, $\chi_1$ particles can be 
produced by these collisions, and some fraction of these $\chi_1$ particles 
subsequently decay within the detector, producing photons that can then be
detected.  At electron beam-dump experiments, $\chi_1$ particles are produced
predominately via bremsstrahlung-like processes such as 
$e^- N \rightarrow e^- N \chi_1 \bar{\chi}_0$, where $N$ denotes an 
atomic nucleus.  At proton beam dumps, these particles are produced
primarily via the decay of various mesons produced by the collisions 
between the incident protons and the nuclei in the target.  
We shall consider the constraints from these two classes of fixed-target
experiments in turn.

Among electron beam-dump experiments, E137~\cite{Bjorken:1988as} and 
E141~\cite{Riordan:1987aw} are the most relevant for
constraining the region of model-parameter space pertinent for FASER.
At each of these experiments, electrons with lab-frame energies 
$E_{e} \sim\mathcal{O}(10)$~GeV collided with a dense target --- aluminum 
in the case of E137, tungsten in the case of E141.  These experiments 
also had similar, $\mathcal{O}({\rm GeV})$ detection thresholds.
By contrast, the beam-dump experiment at Orsay~\cite{Davier:1989wz}, 
which had a similar detection threshold but a far lower electron-beam energy 
$E_e = 1.6$~GeV, yields far weaker constraints within this same region
of parameter space.  
The beam-dump experiment at KEK~\cite{Konaka:1986cb} had an electron-beam energy 
$E_e = 2.5$~GeV, which is also much lower than that of either E137 or E141.  
However, this experiment also had a much lower detection threshold --- around 
$\mathcal{O}(100~{\rm MeV})$.  Nevertheless, given the significant 
background from photon bremsstrahlung at energies $E_\gamma \lesssim 1$~GeV,
this lower detection threshold is unlikely to result in a significant improvement 
in sensitivity to monophoton signals. 
Moreover, the total number of electron collisions attained over the lifetime 
of the KEK experiment is a factor of $10^3$ smaller than the number attained 
at E137.  As a result, we do not expect the KEK experiment to yield
constraints competitive with those from E137 within our 
parameter-space region of interest.

We assess the sensitivity of E137 and E141 within the parameter space
of our inelastic-dark-matter model in a manner analogous to that employed 
in similar studies of other LLP scenarios in
Refs.~\cite{Riordan:1987aw,Andreas:2012mt,Batell:2014mga}.  In particular, 
we simulate the production of photons at these experiments using the 
{\tt MG5\_aMC@NLO} code package~\cite{Alwall:2014hca} and incorporate the 
appropriate form factors for the target material in each experiment --- 
form factors that account for the effects of coherent scattering as well
as nuclear and atomic screening~\cite{Batell:2014mga,Bjorken:2009mm,Kim:1973he}.
We find that neither E137 nor E141 is capable of achieving a statistically 
significant number of events necessary to constrain the region of 
model-parameter space relevant for FASER, even for zero background.  
This is primarily due to the fact that kinematic considerations place 
an upper bound $E_\gamma \leq [E_{e}+(E_{e}^2-m_0^2)^{1/2}] \Delta$ on
the lab-frame energy of a photon produced by $\chi_1\rightarrow \chi_0\gamma$ 
decay at fixed-target experiments of this sort.  Since $\Delta \ll 0.1$
within our parameter-space region of interest, $E_\gamma$ typically lies 
significantly below the threshold for detection.  Indeed, it is only for 
$\Delta > 0.1$ that the E137 and E141 data become constraining for 
inelastic-dipole dark matter.

Among proton beam-dump experiments, Nu-Cal~\cite{Blumlein:2011mv,Blumlein:2013cua}
and CHARM-II~\cite{CHARM-II:1994dzw,Gninenko:2012eq} are the most 
relevant within our parameter-space region of interest.
The Nu-Cal experiment, which primarily functioned as a neutrino detector, 
collided approximately $10^{18}$ protons --- each with a lab-frame energy of 
$E_p \approx 69$~GeV --- with an iron target over the course of its operation.  
Scalar and vector mesons produced in these collisions could decay into $\chi_1$ 
particles via the processes discussed in Sect.~\ref{sec:chiproduction}.~  Decays
of these $\chi_1$ particles to photons within the Nu-Cal detector could then 
give rise to a signal.

In assessing the expected number of signal events at Nu-Cal in 
inelastic-dipole-dark-matter models, we adopt the procedure for
modeling the scalar- and vector-meson production cross-sections 
used in Ref.~\cite{Blumlein:2013cua}.  We relate the inclusive 
cross-section $\sigma\left(p{\rm{Fe}}\rightarrow\pi^0~X\right)$ for
$\pi^0$ production via proton scattering with an iron nucleus $N$
to the inclusive cross-section for $\pi^0$ production via proton-proton
scattering
\be
  \sigma(p N \rightarrow\pi^0 X)~=~A^{\rm{\alpha}}\,
  \sigma(pp\rightarrow\pi^0 X)~,
  \label{eq:ScalePionByNucXSec}
\ee
where $X$ denotes any additional SM particles in the final state, 
where $A=$56 is the atomic mass number of iron, and where $\alpha = 0.55$.
We also model the differential cross-section for $\pi^0$ production
via proton-proton scattering as an average of the corresponding differential
cross-sections for $\pi^+$ and $\pi^-$ production, which have been 
measured empirically~\cite{French-Soviet:1976tin}.
In other words, we take
\begin{eqnarray} 
  \frac{d^2\sigma(pp\rightarrow\pi^0 X)}{dz d(p_T^2)} &~=~&
  \frac{1}{2}\Bigg[\frac{d^2\sigma(pp\rightarrow\pi^-X)}
  {dz d(p_T^2)} \nonumber \\ & & ~~~~~~+
  \frac{d^2\sigma(pp\rightarrow\pi^+ X)}{dz d(p_T^2)}\Bigg]~,~~~~~~~~~
\end{eqnarray}
where $z$ is the fraction of the outgoing momentum of the incident proton
carried by the $\pi^0$ and where $p_T$ is the magnitude of the component of 
its momentum vector transverse to the beam axis. 
Finally, following Ref.~\cite{Chu:2020ysb}, we estimate the differential 
cross-sections for the production all heavier neutral scalar- and vector-meson 
species $M$ as
\begin{equation}
  \frac{d^2\sigma(pN\rightarrow M X)}{dz d(p_T^2)} ~=~
  \mathcal{N}_M^{({\rm POT})}\,
  \frac{d^2\sigma(pN\rightarrow\pi^0 X)}{dz d(p_T^2)}~,
  \label{eq:RescaleDiffXSecMeson}
\end{equation}
where the scaling factor $\mathcal{N}_M^{({\rm POT})}$ represents the average 
number of mesons of that species produced per proton on target.  Conservative
estimates for the $\mathcal{N}_M^{({\rm POT})}$ values for all relevant meson
species were computed in Ref.~\cite{Chu:2020ysb} using
\texttt{Pythia 8.2}~\cite{Sjostrand:2006za,Sjostrand:2014zea} simulations,
and we adopt these $\mathcal{N}_M^{({\rm POT})}$ values as well. 

Given these differential cross-sections, we may calculate the spectrum of 
$\chi_1$ particles produced in the direction of the Nu-Cal detector 
by the decays of these mesons in a straightforward manner.
This detector consisted of a cylindrical decay volume 23~m in length and
2.3~m in diameter located 64~m behind the target with its axis aligned with
the axis of the proton beam.  In assessing the total expected number of signal 
events from $\chi_1\rightarrow \chi_0\gamma$ decays at Nu-Cal for any 
combination of $\Lamcalo$, $m_0$, and $\Delta$, we require not only that the 
$\chi_1$ decay within this decay volume, but also that the lab-frame energy 
of the resulting photon satisfy $E_\gamma > 3$~GeV --- a criterion that derives
from the fact that Nu-Cal was only capable of distinguishing an electromagnetic 
from a hadronic shower when the total energy associated with the shower exceeds 
this threshold.  

By comparing the expected number of signal events for different combinations
of $\Lamcalo$, $m_0$, and $\Delta$ to the number of events satisfying the same 
criteria which were actually observed 
at Nu-Cal during its full run --- a total of 5 such events were observed, and
the expected background was 3.5 events --- we can derive constraints on our 
model-parameter space.  These constraints exclude the shaded region of 
model-parameter space labeled ``$\nu$-Cal'' in the bottom two panels of 
\figref{resultsplots}, which correspond to the parameter choice $\Delta = 0.05$.
Similar exclusion regions do not appear in the other panels of the figure, 
however, because too few photons satisfy the $E_\gamma > 3$~GeV criterion 
for such small values of $\Delta$.  Since the vector mesons $\rho$ and $\omega$, 
which have the highest production rates at Nu-Cal, are kinematically forbidden 
from decaying to $\chi_i$ pairs for $m_0 \gtrsim 390$~MeV, the results of
this experiment have little constraining power within our parameter space
for $m_0$ above this value.

The CHARM experiment and its successor 
CHARM-II~\cite{CHARM-II:1994dzw,Gninenko:2012eq} would likewise have been 
able to produce $\chi_i$ pairs through meson decay.  Thus, the results of
these experiments also constrain the parameter space of 
inelastic-dipole-dark-matter models.
The original CHARM experiment collided approximately $7.1\times10^{18}$ 
protons --- each with an energy of $E_p = 450$~GeV --- with a copper 
target over the course of its operation.  The center of the detector, which was situated
487.3~m behind the target, had a transverse area of $2.4\times2.4~{\rm m}^2$.
Following ~\cite{deNiverville:2018hrc}, which searched for a monophoton 
signal, we take the fiducial decay volume to be 
$2.4 ~\times ~ 2.4 ~\times ~ 11.2~ {\rm m}^3$. 
The CHARM~II experiment collided approximately $2.5\times10^{19}$ 
protons --- each also with an energy of $E_p = 450$~GeV --- with a beryllium 
target over the course of its operation.  The detector, which was situated
870~m behind the target, had a transverse area of $3.7\times3.7~{\rm m}^2$. 
We take the length of the fiducial decay volume to be $35.7$~m, which 
represents the length of the entire calorimeter, ~\cite{CHARM-II:1989nic} 
in the direction parallel to the beam axis.
While the distance between the target and the detector at each of 
these experiments was similar to the distance between FASER and the ATLAS IP, 
the characteristic lab-frame energy of a $\chi_1$ particle produced
by collisions of the proton beams at CHARM and CHARM-II with their 
respective fixed targets was significantly lower than the characteristic 
lab-frame energy of a $\chi_1$ particle produced in the forward direction 
at the LHC.~  As a result, CHARM and CHARM-II data constrain larger values 
of $\Lamcalo$ than does FASER data, which lead to shorter proper lifetimes for 
$\chi_1$.

In deriving the constraints from CHARM data on our parameter space,
we largely follow the procedure outlined in 
Refs.~\cite{deNiverville:2018hrc,Chu:2020ysb}.  Likewise,
in assessing the corresponding constraints from CHARM-II data, we 
largely follow the procedure outlined in Ref.~\cite{Chu:2020ysb}. 
For each experiment, we derive a differential cross-section for $\pi^0$ 
production of the Bonesini-Marchionni-Pietropaolo-Tabarelli-de-Fatis (BMPT)
form~\cite{Bonesini:2001iz} using the 
{\tt{BdNMC}} code package~\cite{deNiverville:2016rqh}.   
We then obtain the differential cross-sections for the production of
$\eta$, $\eta'$, $\rho$, $\omega$, $\phi$ and $J/\psi$ by scaling this
differential cross-section for $\pi^0$ production by an overall factor
$\mathcal{N}_M^{({\rm POT})}$, as in Eq.~(\ref{eq:RescaleDiffXSecMeson}).
 
From these differential cross-sections, we may derive an expected number 
of signal events from $\chi_1$ decays within the decay volume at either
CHARM or CHARM-II for a given combination of $\Lamcalo$, $m_0$, and $\Delta$
in much the same way as we derived the corresponding signal-event count at Nu-Cal.
In accord with the cuts performed on the energy of the electromagnetic 
shower in Ref.~\cite{CHARM:1983ayi}, we require that the lab-frame energy
of the photon satisfy $7.5~{\rm GeV} \leq E_\gamma \leq 50 $~GeV for the 
CHARM analysis.  Likewise, for the CHARM-II analysis, we require that 
$3~{\rm GeV} \leq E_\gamma \leq 24$~GeV in accord with the corresponding cuts
performed in Ref.~\cite{CHARM-II:1994dzw,CHARM-II:1989nic}.  Although the minimum value of 
$E_\gamma$ for both of these windows is considerably higher than the 
threshold value employed in the Nu-Cal analysis above, $E_p$ is
also considerably higher for both CHARM and CHARM-II than it is for Nu-Cal.
As a result, the characteristic energies of the $\chi_1$ particles are 
also significantly higher, and a non-negligible fraction of photons produced
by $\chi_1\rightarrow \chi_0\gamma$ decay survive these $E_\gamma$ cuts,
even for $\Delta = 0.01$.

We derive constraints on our model-parameter space by comparing 
the expected number of signal events obtained for different combinations
of $\Lamcalo$, $m_0$, and $\Delta$ to the number of events 
that were actually observed at CHARM or CHARM-II.
However, there are significant uncertainties in the expected backgrounds at these 
experiments~\cite{deNiverville:2018hrc,CHARM:1983ayi,CHARM-II:1994dzw}.
Thus, since our primary aim in this paper is to identify regions of
our parameter space within which FASER and FASER2 are sensitive, but
which are clearly {\it not}\/ constrained by other experiments, we 
maximize the size of the exclusion regions from CHARM and CHARM-II
by assuming a background of zero events at each experiment. 
The resulting constraint from CHARM-II data excludes the shaded region of 
model-parameter space labeled ``CHARM-II'' in each panel of \figref{resultsplots}.
Since we find that these regions completely subsume those excluded by CHARM data, 
we do not include separate exclusion contours for CHARM in this figure.

\subsection{Constraints from the Elastic Limit\label{sec:elasticLimit}}

In addition to the signals discussed above, which arise as a consequence
of $\chi_1$ decay and therefore vanish in the elastic limit, \ie, the
limit in which $\Delta\rightarrow 0$, there are a number of 
potential signatures of dipole-interacting dark matter 
at colliders, fixed-target experiments, and neutrino facilities that
are associated with processes that do not vanish in that limit.
These signatures include, for example, electron or nucleon recoils
precipitated by collisions of the $\chi_i$ with the detector material 
in fixed-target experiments and monojet~+~$\slashed{E}$ signatures 
at hadron colliders.  The non-observation of such signatures at 
experiments including E613~\cite{Mohanty:2015koa}, 
MiniBooNE~\cite{MiniBooNE:2008paa}, LSND~\cite{LSND:1996jxj}, 
LEP~\cite{Fortin:2011hv}, CHARM-II~\cite{CHARM-II:1994dzw}, and 
LSND~\cite{LSND:1996jxj} places non-trivial constraints on 
inelastic-dipole-dark-matter models.

Since the mass-splittings $m_1 - m_0$ within our parameter-space region of
interest are quite small in comparison with the characteristic energy scales
of these experiments, the corresponding constraints are well approximated 
by those obtained for a purely elastic dipole interaction.  Detailed studies
of the experimental constraints on elastic dipole-interacting dark matter 
have been performed
for both the MDM and EDM cases~\cite{Chu:2018qrm,Chu:2020ysb,Kling:2022ykt}.
Throughout most of our parameter-space region of interest, the leading 
such constraints are those from limits on electron recoils at  
CHARM-II or from limits on monophoton~+~$\slashed{E}$ searches at LEP.~
The shaded region of model-parameter space labeled 
``CHARM-II + LEP ($\Delta$-Ins)'' in each panel of \figref{resultsplots}
is excluded by constraints of this sort.  These constraints collectively 
impose a bound $\Lamcalo \gtrsim 1$~TeV within our parameter-space 
region of interest that is not only independent of $\Delta$, but also 
not particularly sensitive to the value of $m_0$.

\subsection{Astrophysical Constraints\label{sec:AstroConstraints}}

In cases in which $\chi_0$ particles have a significant present-day abundance,
their annihilation within the halos of galaxies via the $t$-channel process 
$\chi_0\overline{\chi}_0\rightarrow\gamma\gamma$ can give rise to a 
monochromatic photon signal at gamma-ray telescopes.  The leading constraints 
on such a signal are currently those from Fermi-LAT data, which imply an 
upper bound on the corresponding thermally-averaged annihilation cross-section 
that varies with $m_0$ from 
$\langle\sigma v\rangle \lesssim 10^{-34}~\rm{cm}^3\rm{s}^{-1}$ for 
$m_0 \approx 1$~MeV to 
$\langle\sigma v\rangle \lesssim 10^{-29}~\rm{cm}^3\rm{s}^{-1}$ for 
$m_0 \approx 1$~GeV~\cite{Bartels:2017dpb}.
This thermally-averaged cross-section  may be estimated as
\be
  \langle\sigma v\rangle  ~\approx~
  \! 10^{-33}~{\rm{cm}}^3\, {\rm{s}}^{-1}
  \! \left(\frac{10^{3}\, {\rm{GeV}}^{}}{\Lamcalo}\right)^4
  \! \left(\frac{m_0}{{\rm{GeV}}}\right)^2 \! \left(\frac{v}{10^{-3}}\right)~,
\ee
which is well below the Fermi-LAT bound for all $m_0$ within our parameter-space
region of interest.  Constraints from other instruments (such as EGRET)  
on monochromatic gamma-ray signals from the annihilation of sub-GeV dark-matter 
particles have also been assessed~\cite{Boddy:2015efa}, but turn out to be 
subleading in comparison with the Fermi-LAT bounds.  

In principle, an additional constraint on inelastic-dipole-dark-matter models
arises from observational limits on distortions of the cosmic microwave
background~\cite{Planck:2015fie}
due to annihilation, co-annihilation, or scattering processes involving the 
$\chi_i$ during or after recombination.  However, we find that these limits do not 
constrain any portion of our parameter-space region of interest that 
is not excluded by other constraints.  Indeed, since the cross-section for the 
annihilation process $\chi_0 \bar{\chi}_0 \rightarrow \gamma \gamma$ is 
proportional to $\Lamcalo^{-4}$, the corresponding annihilation rate is highly 
suppressed for values $\Lamcalo \gtrsim 1$~TeV that are consistent 
with constraints from collider and fixed-target experiments.  Moreover,  
the lifetime of the $\chi_1$ is sufficiently short throughout our parameter-space 
region of interest that the abundance of $\chi_1$ particles is negligible at 
the time of recombination.  At the same time, for $m_0 \gtrsim 1$~MeV and
$\Delta \gtrsim 0.001$, the mass-splitting between $\chi_0$ and $\chi_1$ 
satisfies $m_1 - m_0 \gtrsim 1$~keV.~  Since this significantly exceeds the
temperature $T\sim \cal{O}$(10{\rm ~eV}) during the recombination epoch, 
the production of $\chi_1$ by up-scattering processes immediately prior to 
or during recombination is highly Boltzmann-suppressed.
Taken together, these considerations imply that the co-annihilation rate
during recombination is negligible.

Limits on the effective number $N_{{\rm{eff}}}$ of neutrino 
species during the nucleosynthesis epoch place stringent 
constraints~\cite{Sabti:2019mhn} on light thermal relics with highly 
suppressed couplings to the visible sector.
However, such constraints only become relevant for $m_0\lesssim 10$~MeV,
and do not impact the region of model-parameter space relevant for FASER.

Finally, direct-detection experiments also place constraints on the
scattering cross-sections of $\chi_0$ with atomic nuclei or with electrons.
However, inelastic-scattering rates for a dark-matter particles with 
$m_0 \sim \mathcal{O}({\rm GeV})$ at such experiments are highly suppressed
for $\Delta \gtrsim 10^{-6}$.  While elastic scattering via loop-level processes
still yield a contribution to the overall scattering rate for larger values of 
$\Delta$~\cite{Berlin:2018jbm}, this contribution is far too small within our
parameter-space region of interest to be constraining.


\section{Conclusions and Directions for Future Research\label{sec:Conclusion}}


In this paper, we investigated the prospects for detecting a signature
of inelastic-dipole dark matter at the FASER detector and its
proposed upgrade FASER2.  This signature involves the production of the 
heavier of the two dark-sector states present in the theory by collisions at the 
ATLAS IP and its subsequent decay into the lighter dark-sector state and a 
single photon inside the FASER decay volume.  We determined the 
discovery reach of FASER and FASER2 within the parameter space of this 
model for the case of either a magnetic or an electric dipole-moment
interaction and showed that these detectors are capable of probing 
regions of this parameter space within which other experimental probes have
little or no sensitivity.  Moreover, these include regions wherein thermal
freeze-out yields an abundance for the lighter dark-sector state that
agrees with the observed abundance of dark matter.

Several comments are in order.  First, 
it is worth noting that one of the reasons why FASER and FASER2 are 
capable of probing regions of inelastic-dipole-dark-matter parameter space 
within which existing experiments have little constraining power is simply 
the far higher energy of the LHC.
Indeed, as we saw in Sect.~\ref{sec:ExistingConstraints}, one of the 
reasons why current fixed-target bounds are not terribly constraining
within our parameter-space region of interest is that the majority
of photons produced by $\chi_1$ decays within the detector at these
experiments would have had lab-frame energies that are below the
corresponding detection thresholds.  By contrast, the substantial boost
provided to $\chi_1$ particles at the LHC ensures that the photons 
produced by the decays of these particles within the FASER detector have
far higher energies.  Moreover, the time-dilation factor provided by 
these large boosts enhances the ability of FASER to probe regions of 
parameter space in which the lifetime of $\chi_1$ is considerably shorter.
Since small mass splittings are a generic possibility that leads 
to long-lived particles, and since these small mass splittings in turn 
lead to soft decay products, one can expect the high energies and boosts 
available at the LHC to be advantageous for probing many similar 
LLP models beyond the specific one discussed here. 
Such models might include, for example, inelastic-dark-matter models 
involving other coupling structures, such as the anapole or charge-radius 
operators ~\cite{Chu:2018qrm,Chu:2020ysb,Kling:2022ykt}.

In this connection, it is worth noting that other proposed experiments
would also be able to probe parts of our region of interest within the 
parameter space of inelastic-dipole dark matter models.
At CERN, these include the SHiP experiment~\cite{Bonivento:2013jag,SHiP:2015vad},  
which would collide a 400~GeV proton beam with a fixed target.  An 
estimate of the reach of this experiment within our parameter-space 
region of interest is provided in Appendix~\ref{app:SHiP}.  As can be seen 
there, the reach depends critically on the mass splitting between 
the dark states and the energy threshold for monophoton signal detection.  
For small $\Delta$, FASER and FASER2 probe regions beyond the reach of SHiP, 
but for larger $\Delta$, their reaches may be very roughly comparable.  Further detailed
studies of the energy threshold for monophoton signal detection are clearly needed. In Ref.~\cite{Barducci:2022gdv}, the authors explored the sensitivity of SHiP and FASER to an inelastic sterile neutrino interacting through an MDM operator and found SHiP to be more constraining than FASER. This is a consequence of their requiring very high thresholds for monophoton detection for FASER  ($E_{\gamma} \ge 10 ~{\rm GeV}$), while requiring much lower thresholds at SHiP  ($E_{\gamma} \ge 0.1 ~{\rm GeV}$). Although these thresholds are not precisely known for each experiment, FASER has significantly more shielding against background and is likely better suited to finding very low-energy signal photons.
We also note that other proposed experiments, such as SHADOWS~\cite{Baldini:2021hfw}
have the potential to probe this parameter-space region, although the same comments about energy thresholds apply for all fixed target experiments. Other LLP detectors such as  ANUBIS~\cite{Bauer:2019vqk}, CODEX-b ~\cite{Aielli:2019ivi}, MAPP~\cite{Acharya:2022nik}, MATHUSLA~\cite{MATHUSLA:2019qpy} are primarily designed to be sensitive to LLPs produced with  significantly more transverse momenta than those considered here. Thus data from these detectors is not expected to be competitive with data from FASER in the parameter space we consider.

Finally, in this paper we have considered the 
case of a dark sector that comprises two states $\chi_0$ and $\chi_1$ that
interact with each other and with the visible sector via an 
inelastic-dipole interaction.  While this model is interesting in its own right,  
it can also be viewed as merely the simplest possible realization of a 
non-minimal dark sector consisting of $N$ different states $\chi_i$ with 
$i = 0, 1, \ldots, N - 1$ that interact with each other via a set of operators 
$\mathcal{O}_{ij}$, each with the MDM or EDM structure given 
in Eq.~(\ref{eq:lagrangianEM}).  Within this more general class of
models, decay cascades involving multiple sequential LLP decays of the form 
$\chi_i\rightarrow \chi_j \gamma$ can develop.  Scenarios in which decay  
cascades involving multiple LLPs can arise frequently give rise to distinctive 
phenomenological signatures~\cite{Dienes:2021cxr} that differ from those 
observed in more traditional LLP scenarios.  It would therefore be 
interesting to explore the prospects for probing dark-sector scenarios of 
this sort at FASER, FASER2, and other proposed experiments.


\section*{Acknowledgements}


We thank Daniele Barducci, Enrico Bertuzzo, David Casper, Xiaoyong Chu, Patrick deNiverville, Andrei Golutvin, 
Felix Kling, Dominik K\"{o}hler, Jui-Lin Kuo, Gopolang Mohlabeng, Marco Taoso, Claudio Toni and Sebastian 
Trojanowski for discussions.
The research activities of K.R.D.\ were supported in part by the U.S.\ Department of Energy 
under Grant DE-FG02-13ER41976 / DE-SC0009913, and by the U.S.\ National Science 
Foundation (NSF) through its employee IR/D program.  The work of J.L.F., M.F., 
and F.H.\ was supported in part by NSF Grants PHY-1915005 and PHY-2210283. The work 
of J.L.F.\ was also supported in part by NSF Grant PHY-2111427, Simons Investigator 
Award \#376204, Simons Foundation Grant 623683, and Heising-Simons Foundation 
Grants 2019-1179 and 2020-1840. The work of M.F. was also supported in part by NSF Graduate Research Fellowship
Award No. DGE-1839285. The work of F.H.\ was also supported in part by 
the International Postdoctoral Exchange Fellowship Program. 
The research activities of S.L.\ were supported in part by the National Research 
Foundation of Korea (NRF) grant funded by the Korean government (MEST) 
(No. NRF-2021R1A2C1005615) and the Samsung Science \& Technology Foundation 
under Project Number SSTF-BA2201-06.  The research activities 
of B.T.\ were supported in part by the National Science Foundation under Grant PHY-2014104.
The opinions and conclusions expressed herein are those of the authors and do not represent 
any funding agencies.


\appendix



\section{Meson Branching Fractions \label{app:BFmesons}}


In this Appendix, we provide explicit formulas for the branching fractions
for the meson-decays processes that contribute to $\chi_1$ production in 
an inelastic-dipole-dark-matter context.

The dominant contribution to $\chi_1$ production from each relevant neutral 
vector meson $V$ is that associated with the two-body decay process 
$V\rightarrow \overline{\chi}_0 \chi_1$.  It is convenient to parametrize 
the branching fraction for each such decay process in terms of the 
corresponding branching fraction for the SM decay process 
$V\rightarrow e^+e^-$ as follows:
\begin{eqnarray}
  && {\rm BR}(V \to \overline{\chi}_0 \chi_1)
    ~=~ {\rm BR}(V \to e^+e^-) \,\frac{\Gamma_{V \to \overline{\chi}_0\chi_1}}
    {\Gamma_{V \to e^+e^-}} \nonumber \\
  && ~~~~~=~ {\rm BR}(V \to  e^+e^-) 
    \,\frac{|{\cal M}(V \to \overline{\chi}_0\chi_1)|^2 }
    {   |{\cal M}(V \to  e^+e^-)|^2} \,
    \frac{\abs{\vec{\mathbf{p}}_{\chi_0}}}{\abs{\vec{\mathbf{p}}_e}} ~,~~~~~~~
\end{eqnarray}
where $|{\cal M}(V \to \overline{\chi}_0\chi_1)|^2$ and 
$|{\cal M}(V \to  e^+e^-)|^2$ are the squared matrix elements for the
two decay processes, averaged over initial-state-particle spins and 
summed over final-state particle spins, and where $|\vec{\mathbf{p}}_{\chi_0}|$
and $|\vec{\mathbf{p}}_e|$ denote the magnitudes of the three-momenta 
of either final-state particle in the corresponding process.  

The squared matrix element for vector-meson decay to electrons is given by
\be
  |{\cal M}(V \to  e^+e^-)|^2
  ~=~ \frac{16\pi \alpha}{3}\, m^2_V\,
  \left( 1+\frac{2m_e^2}{m_V^2} \right)~.
\ee
For the case of an MDM operator structure, the squared matrix element for the 
process $V \to \overline{\chi}_0 \chi_1$ is given by
\begin{eqnarray}
  && \left| {\cal M}_{{\rm MDM}} (V \to \overline{\chi}_0 \chi_1) \right|^2  
    ~=~ \frac{8 m_V^4}{3   \Lambda_m^{2}} \nonumber \\  
    && ~~~~~ \times
    \left[1+\frac{m_0^2+6m_0m_1+m_1^2}{m_V^2}-
    \frac{2(m_1^2-m_0^2)^2}{m_V^4}\right]~.~~~~~~~~
    \label{eq:MDMamplitude}
\end{eqnarray}
For the case of an MDM operator structure, the squared matrix element for this
same process is given by
\begin{eqnarray}
  && | {\cal M}_{{\rm EDM}} (V \to \overline{\chi}_0 \chi_1) |^2 
     ~=~  \frac{8 m_V^4}{3 \Lambda_e^{2}}  \nonumber \\ 
  && ~~~~~ \times
    \left[ 1+\frac{2(m_1-m_0)^2}{m_V^2}\right] 
    \left[ 1-\frac{(m_1+m_0)^2}{m_V^2}\ \right] ~.~~~~~~~
    \label{eq:EDMamplitude}
\end{eqnarray}
We note that for $m_1=m_0$, the expressions in Eqs.~(\ref{eq:MDMamplitude}) 
and~(\ref{eq:EDMamplitude}) reduce to the corresponding results obtained in 
Ref.~\cite{Chu:2020ysb} for the case of a purely elastic dipole interaction.  

By contrast, the dominant contribution to $\chi_1$ production from each relevant
neutral pseudoscalar meson $P$ arises is that associated with the three-body decay 
process $P\rightarrow \gamma^\ast \gamma\rightarrow \overline{\chi}_0 \chi_1\gamma$, 
where $\gamma^\ast$ denotes an off-shell photon.  The coupling between each 
$P$ and a pair of photons is described by the effective 
Lagrangian 
\begin{equation}
  {\cal L}_{P} ~=~ \frac{e^2}{16\pi^2 F_{P}} \,
    P F_{\mu \nu}\widetilde{F}^{\mu\nu}~, 
\end{equation}
where $F_P$ is the corresponding pseudoscalar decay constant and where
$\widetilde{F}^{\mu\nu} \equiv \epsilon^{\mu\nu\rho\sigma}F_{\rho\sigma}/2$ 
is the dual electromagnetic field-strength tensor. 

The differential branching fraction with respect to the energy $E_1^*$ of the
final-state $\chi_1$ particle in the rest frame of the decaying $P$ can be 
written in the form
\begin{equation}
\frac{d{\rm BR(P\rightarrow  \overline{\chi}_0\chi_1 \gamma)}}{dE_1^*} ~=~ 
  \frac{\rm BR(P\rightarrow \gamma \gamma)}{\Gamma({\rm P\rightarrow \gamma \gamma})}
  \times \frac{d\Gamma(P\rightarrow\overline{\chi}_0\chi_1\gamma)}{dE_1^*}~,~
    \label{eq:dBRdEstar}
\end{equation}
where $d\Gamma(P\rightarrow\overline{\chi}_0\chi_1\gamma)/dE_1^*$ denotes the 
corresponding differential partial decay width and where
\begin{equation}
  \Gamma(P\rightarrow \gamma \gamma) ~=~ \frac{\alpha^2 m_{\pi}^3}{32\pi^3F_P^2}
\end{equation}
denotes the full partial width of $P$ to photon pairs.  This differential partial
width is given by
\begin{eqnarray}
  \frac{d\Gamma(P\rightarrow\overline{\chi}_0\chi_1\gamma)}{dE_1^*} &\,=\,&
    \int \frac{d^3p_0d^3p_{\gamma}d\Omega_1}{16(2\pi)^5m_P}
    |{\cal M} (P \to \overline{\chi}_0 \chi_1\gamma)|^2  
      \nonumber \\ & & ~ \times
     \frac{p_1^*}{E_0E_\gamma}
      \delta^4(p_{\gamma}+p_1+p_0-m_P)~, \nonumber \\
\end{eqnarray}
where $| {\cal M} (P \to \overline{\chi}_0 \chi_1\gamma) |^2$ 
is the squared matrix element for the decay process 
$P\rightarrow\overline{\chi}_0 \chi_1\gamma$, summed over final-state-particle 
spins.  For the case
of an MDM operator structure, this squared matrix element is given by
\begin{eqnarray}    
  && | {\cal M}_{{\rm MDM}} (P \to \overline{\chi}_0 \chi_1\gamma) |^2 ~=~ 
    \frac{8e^4}{\pi^4 F_{P}^2 \Lambda_m^2 q^4}
    \nonumber \\
    &&~~~~ \times \Big[ (2k_0\cdot q)(k_1 \cdot q)
      (k_\gamma\cdot q)^2
      +q^4(k_\gamma\cdot k_0)(k_\gamma\cdot k_1)~~~~~~~
      \nonumber \\
    && ~~~~~~~~ - q^2(k_\gamma\cdot k_1)(k_\gamma\cdot q)(k_0\cdot q)
      -m_0m_1q^2(k_\gamma\cdot q)^2
      \nonumber \\
    &&~~~~~~~~ -q^2(k_\gamma\cdot k_0)(k_\gamma\cdot q)(k_1\cdot q) \Big]~,
\end{eqnarray}
where $p$, $k_\gamma$, $k_0$, and $k_1$ respectively denote the four-momenta 
of $P$, $\gamma$, $\chi_0$, and $\chi_1$, and where $q$ denotes the four-momentum 
of the virtual photon.
By contrast, for the case of an EDM operator structure, this matrix element is given by
\begin{eqnarray}
    && | {\cal M}_{{\rm EDM}} (P \to \overline{\chi}_0 \chi_1\gamma) |^2 ~=~ 
      \frac{8e^4}{\pi^4 F_{P}^2 \Lambda_e^2 q^4}
    \nonumber \\
    && ~~~~ \times \Big[ (2k_0\cdot q)
      (k_1 \cdot q)(k_\gamma\cdot q)^2
      + q^4(k_\gamma\cdot k_0)(k_\gamma\cdot k_1)  ~~~~~~~
    \nonumber \\
    &&~~~~ ~~~~ -  q^2(k_\gamma\cdot k_1)(k_\gamma\cdot q)(k_0\cdot q)
      + m_0m_1q^2(k_\gamma\cdot q)^2
    \nonumber \\
    && ~~~~~~~~ -q^2(k_\gamma\cdot k_0)(k_\gamma\cdot q)(k_1\cdot q) \Big]~. 
\end{eqnarray}
We note that the factors of $F_P$ in $\Gamma(P\rightarrow \gamma \gamma)$
and $d\Gamma(P\rightarrow\overline{\chi}_0\chi_1\gamma)/dE_1^*$ cancel
in Eq.~(\ref{eq:dBRdEstar}), while ${\rm BR(P\rightarrow \gamma \gamma)}$
is well approximated by its SM value 


\section{Thermally-Averaged Cross-Sections and Decay Rates\label{app:XSecs}}


In this Appendix, we provide expressions for the relevant thermally-averaged
quantities that appear in the collision terms Eq.~(\ref{eq:collision}).

In general, the thermally-averaged decay width for a particle that decays via the 
two-body process $a\to b + c$, where the species $c$ and $d$ in the final 
state are in thermal equilibrium with the SM radiation bath, is given by
\beqn
  \expt{\Gamma_{a\to bc}}~=~ \frac{g_a}{n_a^{\rm eq}}\int
  \frac{d^3p_a}{(2\pi)^3}\, \frac{m_a}{E_a}\,e^{-E_a/T}\,
  \Gamma_{a\to bc}~,~~~~~~~
\eeqn
where $g_a$, $n_a^{\rm eq}$, $m_a$, and $\Gamma_{a \to bc}$, respectively denote the
number of internal degrees of freedom, and equilibrium number density, mass, and proper 
decay width of the initial-state particle $a$.  Since the equilibrium number
density is
\begin{equation}
  n_a^{\rm eq} ~=~ g_a\frac{m_a^2T}{2\pi^2}K_2(m_a/T)~,
\end{equation}
where $K_\nu(x)$ denotes the modified Bessel function of the second kind,
one finds that
\begin{equation}
  \expt{\Gamma_{a\to bc}} ~=~ \frac{K_1(m_a/T)}{K_2(m_a/T)}\,
  \Gamma_{a\to bc}~.
\end{equation}

The sole thermally-averaged decay width that appears in 
Eq.~(\ref{eq:collision}) is the one for the decay process
$\chi_1 \rightarrow \chi_0\gamma$.  The corresponding proper decay width 
from which this thermal average is derived is provided in 
Eq.~(\ref{eq:GammaChi1}).

The thermally-averaged cross-section for the $2\to 2$ scattering process of
the form $a+b\to c+d$, where the species $c$ and $d$ in the final state are 
in thermal equilibrium with the SM radiation bath, can be written 
as~\cite{Cannoni:2013bza}
\begin{equation}
  \expt{\sigma v} \,=\, \frac{1}{n_a^{\rm eq}n_b^{\rm eq}}
  \int\!\frac{d^3 p_a}{(2\pi)^3}\frac{d^3 p_b}{(2\pi)^3} \,
  \sigma_{ab\to cd}\, v_{\rm rel}\,
  e^{-(E_a + E_b)/T}~,
  \label{eq:SigmavRaw}
\end{equation}
where $\sigma_{ab\to cd}$ is the invariant cross-section for the scattering 
process and where  
\begin{equation}
  v_{\rm rel} ~\equiv~ \frac{\sqrt{(p_a\cdot p_b)^2-m_a^2 m_b^2}}{p_a\cdot p_b}
\end{equation}
denotes the relative velocity of $a$ and $b$ in the background frame.  
To a very good approximation, when $a$ and $b$ are highly non-relativistic,
this relative velocity coincides with the M{\o}ller velocity
and the expression in Eq.~(\ref{eq:SigmavRaw}) simplifies 
to~\cite{Gondolo:1990dk}
\begin{eqnarray}
  \expt{\sigma v} &\,=\,& 
    \frac{T}{32\pi^4 n_a^{\rm eq}n_b^{\rm eq}}\int_{s_{\rm min}}^\infty \!\! ds \,
    \left(s-s_{\rm min}\right)[s-(m_a-m_b)^2] \nonumber \\
  &&~~~~~~~~~~~~~~~~~~~~~\times \frac{\sigma_{ab\to cd}}{\sqrt{s}}
    K_1 \!  \left(\frac{\sqrt{s}}{T}\right)\,,
\label{eq:thermal_avg}
\end{eqnarray}
where $s_{\rm min} \equiv (m_a + m_b)^2$.  For the case in which $a$ and
$b$ are of the same species, this expression further reduces to
\begin{eqnarray}
  \expt{\sigma v} ~&=&~ \frac{T}{32\pi^4 ({n_a^{\rm eq}})^2}
    \int_{4m_a^2}^\infty ds\, (s-4m_a^2) \nonumber \\
  & & ~~~~~~~~~~~~~~~~~~~~~\times \sqrt{s} \, \sigma_{ab\to cd} 
    K_1\!\left(\frac{\sqrt{s}}{T}\right)~.~~~~~~
\end{eqnarray}

We now turn to discuss the cross-sections for the individual processes 
whose thermal averages appear in Eq.~(\ref{eq:collision}).
For both the MDM and EDM operator structures, the cross-section for the 
$s$-channel co-annihilation process $\overline{\chi}_0 \chi_1 \to e^+ e^-$ 
can be written as
\beq
  \sigma_{01\to ee} ~=~ \frac{e^2}{6\pi\Lamcalo^2}
  \frac{\sqrt{s-4m_e^2}~C^{(s)}}{s^{5/2}
  \sqrt{(s-\smin)[s-(m_0-m_1)^2]}}\,,
  \label{eq:eeSExpr}
\eeq
where $\smin=(m_0+m_1)^2$ and where the form of $C^{(s)}$ 
differs between the MDM and EDM cases.  For the MDM case, this 
quantity takes the form
\begin{equation}
  C^{(s)}~=~ C^{(s)}_1+C^{(s)}_2~,
  \label{eq:CeeSExprMDM}
\end{equation}
where we have defined
\begin{eqnarray}
  C^{(s)}_1 \,&\equiv&\, 2m_e^2[s^2  +(\smin+4m_0 m_1)s-2\smin(m_0  - m_1)^2] \nn\\
  C^{(s)}_2 \,&\equiv&\, s\!\left[s^2 + (\smin +  4m_0 m_1)s  -  
    \frac{7}{2}\smin(m_0 -  m_1)^2\right]~ .\nonumber\\
\end{eqnarray}
By contrast, for the EDM case, $C^{(s)}$ takes the form
\begin{eqnarray}
  C^{(s)} ~ &=&~ (s+2m_e^2) \bigl[ s^2  + (\smin-8m_0 m_1)s ~~~~~~~~~
  \nonumber\\
    &&~~~~ - 2\smin(m_0  - m_1)^2\bigr]~ .~~~
    \label{eq:CeeSExprEDM}
\end{eqnarray}

For processes in which $\overline{\chi}_0$ and $\chi_1$ co-annihilate into hadronic 
final states, we model the inclusive cross-section using the $R(s)$ function given 
in Ref.~\cite{ParticleDataGroup:2022pth}.  In particular, we take
\beq
  \sigma_{01\to \text{hadrons}}~=~\sigma_{01\to\mu\mu}
   \, R(s)\,\Theta\left(s-4m_\pi^2\right)~, 
\eeq
where $\Theta(x)$ denotes the Heaviside theta function and where $m_\pi$ is 
the pion mass.

For both the MDM and EDM operator structures, the cross-section for the 
$t$-channel annihilation process $\overline{\chi}_0 \chi_0  \to \gamma \gamma$ 
can be written as
\begin{eqnarray}
  && \sigma_{00\to\gamma\gamma} ~=~ 
    \frac{2}{\pi\Lamcalo^4 }\, \frac{C^{(t)}}{s(s-4m_0^2)} ~ ,
    \label{eq:chi0chi0gammagamma}
\end{eqnarray}
where the form of $C^{(t)}$ differs between the MDM and EDM cases.  In the  
MDM case, this quantity takes the form
\begin{equation}
  C^{(t)} ~=~ \sum_{\ell=1}^5 C_\ell^{(t)} 
\end{equation}
where we have defined
\begin{eqnarray}
  C^{(t)}_1 ~&\equiv&~  
    \frac{1}{3}\left[\left(m_0^2-X_+\right)^3-\left(m_0^2-X_-\right)^3\right] \nonumber\\
  C^{(t)}_2 ~&\equiv&~  (m_0^2-m_1^2)^4\nn\\
    &~&~~\times\Bigg( \frac{1}{m_1^2-m_0^2+X_+}-\frac{1}{m_1^2-m_0^2+X_-} \Bigg) \nonumber\\
  C^{(t)}_3 ~&\equiv&~ \frac{(s \! + \! 2m_1^2 \! - \! 4m_0^2)}{2} \! 
    \left[\left(m_0^2 \! - \! X_+\right)^2
    \! \! - \! \! \left(m_0^2 \! - \! X_-\right)^2\right]\nn\\
  C^{(\rm t)}_4~&\equiv&~\left[6m_0^4 \! + \! 3m_1^4 \! - \! m_0^2
    \left(3s+8m_1^2\right)\right]\left(X_- \! \! - \! \! X_+\right) \nonumber\\
  C^{(\rm t)}_5~&\equiv&~\frac{1}{s+2\left(m_1^2-m_0^2\right)} \nn \\
  &~&~~\times\bigg[8 \left(m_0^2-m_1^2\right)^4-4(m_0^2-m_1^2)^2(2m_0^2-m_1^2) s\nn\\
  &~&~~~~~~+(m_0^4 - 4m_0^2m_1^2 - m_1^4) s^2 \bigg]\nn\\
  &~&~~\times\log\left(\frac{X_+-m_0^2+m_1^2}{X_--m_0^2+m_1^2}\right)\, ,
\end{eqnarray}
in terms of the quantities
\begin{equation}
  X_\pm ~\equiv~ \frac{1}{2}\left(s\pm \sqrt{s(s-4m_0^2)}\, \right) .
  \label{eq:XpmDef}
\end{equation}
By contrast, for the EDM case, $C^{(t)}$ takes the form
\begin{equation}
  C^{(t)}~=~\frac{1}{s-2(m_0^2-m_1^2)}
    \, \frac{X_+-X_-}{m_1^2 s+(m_0^2-m_1^2)^2}\, 
  \sum_{\ell=1}^6 C_\ell^{(t)}\,,
\end{equation}
where we have defined
\begin{eqnarray}
  C^{(\rm t)}_1 ~&\equiv&~ 8(m_0^2-m_1^2)^5\nonumber\\
  C^{(\rm t)}_2 ~&\equiv&~ -\frac{4}{3}(5m_0^2-6m_1^2)(m_0^2-m_1^2)^3 s\nonumber\\
  C^{(\rm t)}_3 ~&\equiv&~  \frac{1}{3}(3m_0^6-19m_0^4 m_1^2 +15m_0^2 m_1^4 +m_1^6)s^2 \nonumber\\
  C^{(\rm t)}_4 ~&\equiv&~ \frac{1}{6}(m_0^4+4m_0^2m_1^2+9m_1^4)s^3 \nonumber\\
  C^{(\rm t)}_5 ~&\equiv&~\frac{1}{6}m_1^2 s^4\nonumber\\
  C^{(\rm t)}_6 ~&\equiv&~\bigg[8 \left(m_0^2-m_1^2\right)^4-4(m_0^2-m_1^2)^2(2m_0^2-m_1^2) 
    s\nonumber\\
    && \! \! \! \! \! \! \! +(m_0^4 - 4m_0^2m_1^2 - m_1^4) s^2 \bigg] 
    \log\left(\frac{X_+-m_0^2+m_1^2}{X_--m_0^2+m_1^2}\right)\, ,\nonumber\\
\end{eqnarray}
and where $X_\pm$ are once again given in Eq.~(\ref{eq:XpmDef}).
The corresponding expressions for the $t$-channel annihilation process 
$\overline{\chi}_1 \chi_1 \to \gamma \gamma$ in the MDM and EDM cases are identical 
in form to the expression in Eq.~(\ref{eq:chi0chi0gammagamma}), but with 
$m_0 \leftrightarrow m_1$ exchanged in this cross-section formula and all 
ancillary expressions.

For both the MDM and EDM operator structures, the cross-section for the process 
$\chi_0 e^\pm \to \chi_1 e^{\pm}$, in which a $\chi_0$ particle up-scatters into 
a $\chi_1$ particle off an electron or positron, can be written as 
\begin{equation}
  \sigma_{0e\to 1e} ~=~ 
    \frac{e^2 C^{\rm (sc)}}{2\pi\Lamcalo^2 s(s-\smin) \left[s-(m_0-m_e)^2\right]}
  \label{eq:UpscatterOffeXSec}
\end{equation}
where $\smin =(m_0+m_e)^2$ and where the form of $C^{({\rm sc})}$ differs 
between the MDM and EDM cases.  In the MDM case, this quantity takes the form
\begin{eqnarray}
  C^{\rm (sc)}~&=&~C^{(\rm sc)}_1\left[\left(m_0+m_1\right)^2+2m_e^2-4s\right]
    \nonumber \\
  &&~+s\Big[2s^2-4m_e^2(s+m_0m_1)+2m_e^4+m_0^4+m_1^4\nn\\
  &&~~~~-C^{(\rm sc)}_3\Big]\log \! \left(\frac{C^{(\rm sc)}_3 \! - \! C^{(\rm sc)}_1
   \! - \! C^{(\rm sc)}_2}{C^{(\rm sc)}_1 \! - \! C^{(\rm sc)}_2 \! 
   + \! C^{(\rm sc)}_3} \right)~,~~~ 
\end{eqnarray}
where we have defined
\begin{eqnarray}
  C^{(\rm sc)}_1 ~&\equiv&~ \sqrt{(s-\smin)\left[s-(m_e-m_0)^2 \right]} \nonumber \\
    && ~~~\times \sqrt{\left[s \! - \! (m_e+m_1)^2\right]
    \left[s \! - \! (m_e-m_1)^2 \right]} \nonumber\\
  C^{(\rm sc)}_2
    ~&\equiv&~\left(s+m_0^2-m_e^2\right)\left(s+m_1^2-m_e^2\right) \nonumber\\
  C^{(\rm sc)}_3~&\equiv&~2s\left(m_0^2+m_1^2\right)~.
\end{eqnarray}
By contrast, in the EDM case, $C^{\rm (sc)}$ takes the form
\begin{eqnarray}
  C^{\rm (sc)}~&=&~C^{(\rm sc)}_1\left[\left(m_0-m_1\right)^2+2m_e^2-4s\right]
    \nonumber\\
  &&~+s\Big[2s^2-4m_e^2(s-m_0m_1)+2m_e^4+m_0^4+m_1^4\nn\\
  &&~~~-C^{(\rm sc)}_3\Big]\log \! \left(\frac{C^{(\rm sc)}_3 \!- \! C^{(\rm sc)}_1
   \! - \! C^{(\rm sc)}_2}{C^{(\rm sc)}_1 \! - \! C^{(\rm sc)}_2 \! 
   + \! C^{(\rm sc)}_3} \right) \,.~~
\end{eqnarray}
The cross-section for the corresponding down-scattering process 
$\chi_1 e^\pm \to \chi_0 e^\pm$ may be obtained simply by exchanging 
$m_0 \leftrightarrow m_1$ everywhere in Eq.~(\ref{eq:UpscatterOffeXSec}) and 
all ancillary expressions.  Moreover, the corresponding expression for the 
up-scattering of $\chi_0$ into $\chi_1$ off any other electrically charged 
SM particle may be obtained by replacing $m_e$ with the mass of that SM
particle everywhere in Eq.~(\ref{eq:UpscatterOffeXSec}) and 
all ancillary expressions.


\section{Prospects for Detection at SHiP\label{app:SHiP}}


SHiP~\cite{Bonivento:2013jag,SHiP:2015vad} is a proposed experiment at CERN in 
which a $400$-GeV proton beam will collide with a fixed target consisting of 
tungsten and a titanium-zirconium-doped molybdenum alloy. SHiP probes lifetimes 
comparable to FASER and FASER2 and so it is useful to compare the reach of 
these experiments. 


\begin{figure*}
    \centering
    \includegraphics[width=0.49\textwidth]{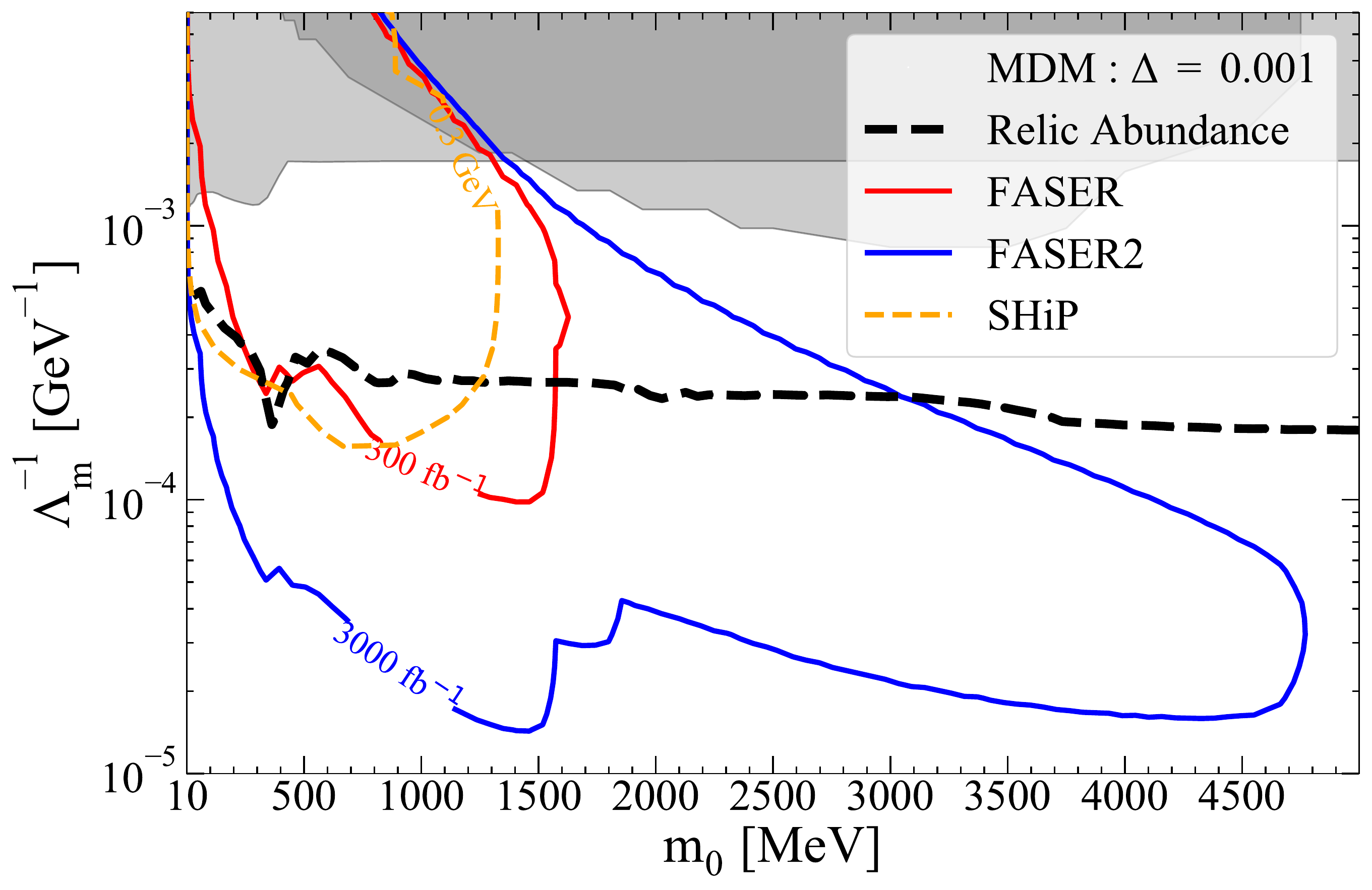}
    \includegraphics[width=0.49\textwidth]{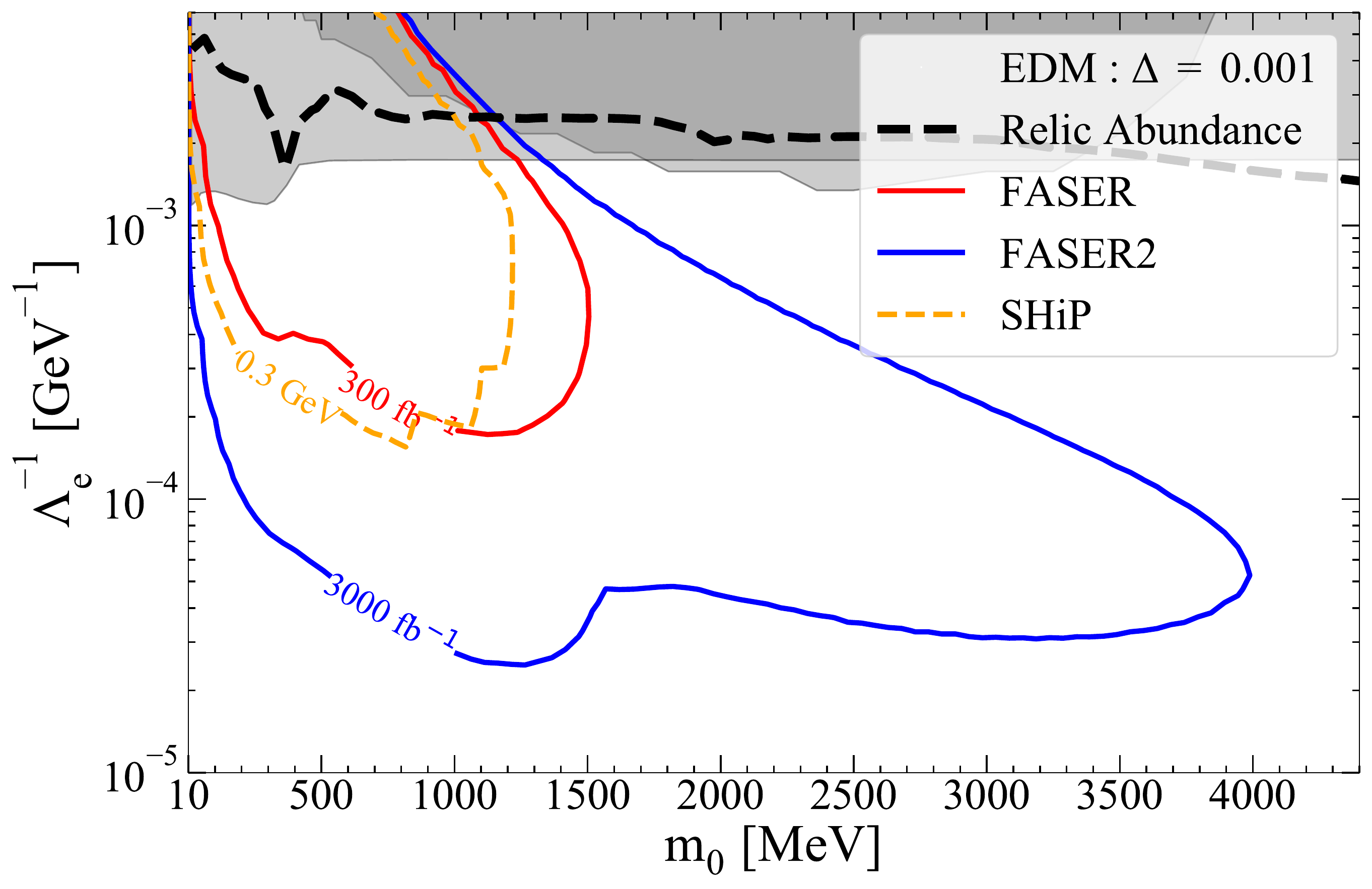}
    \includegraphics[width=0.49\textwidth]{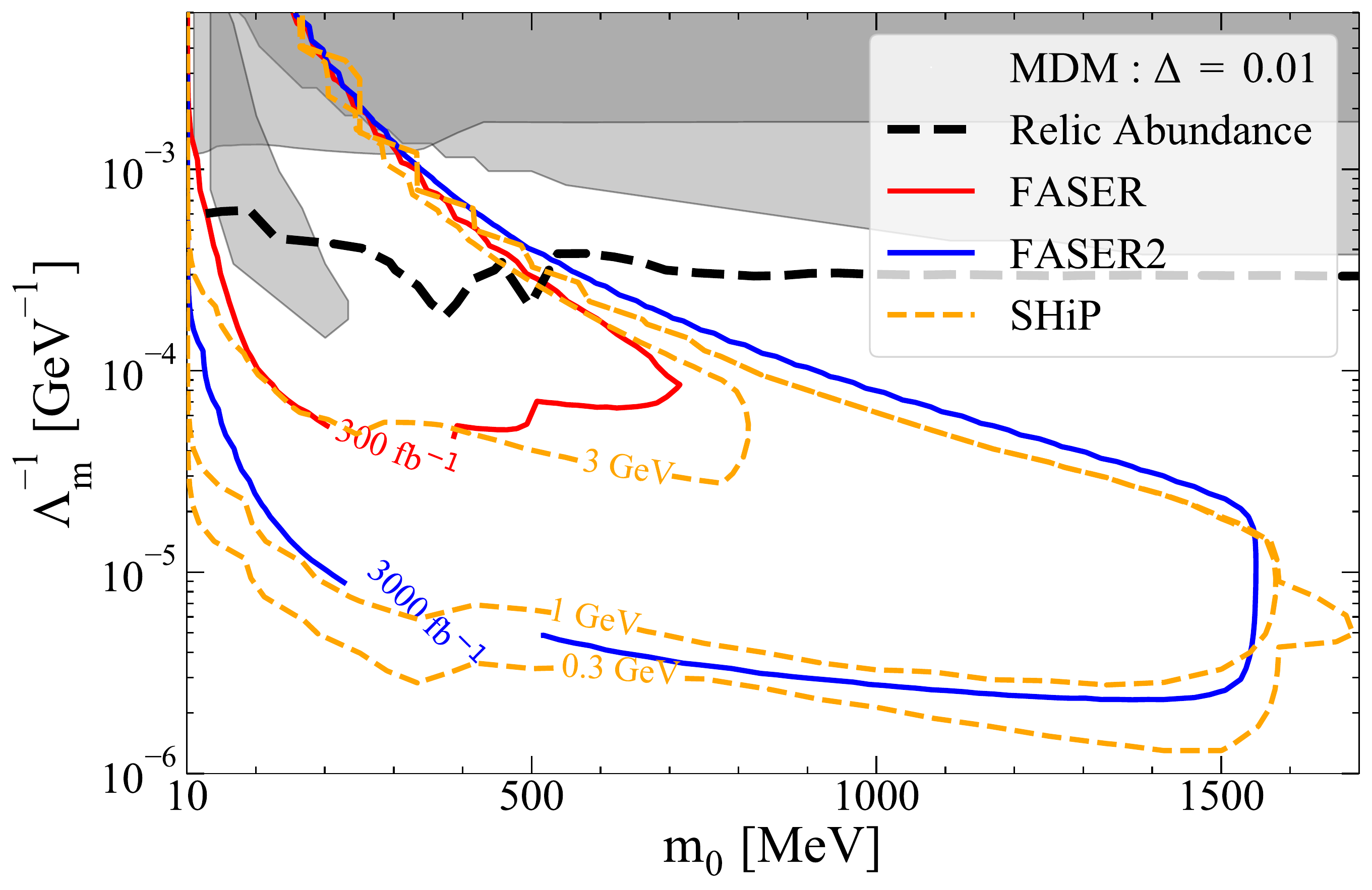}
    \includegraphics[width=0.49\textwidth]{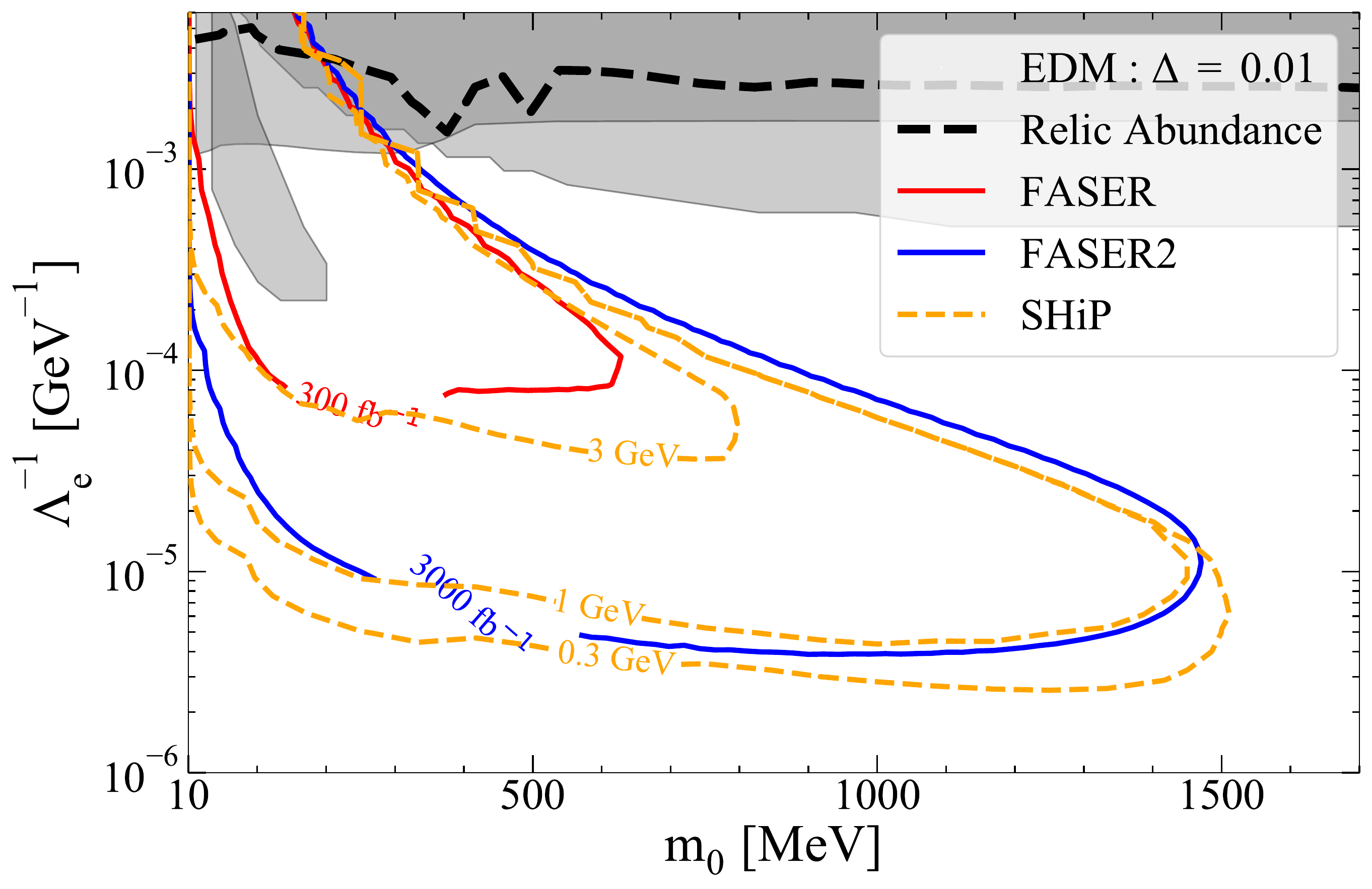}
    \includegraphics[width=0.49\textwidth]{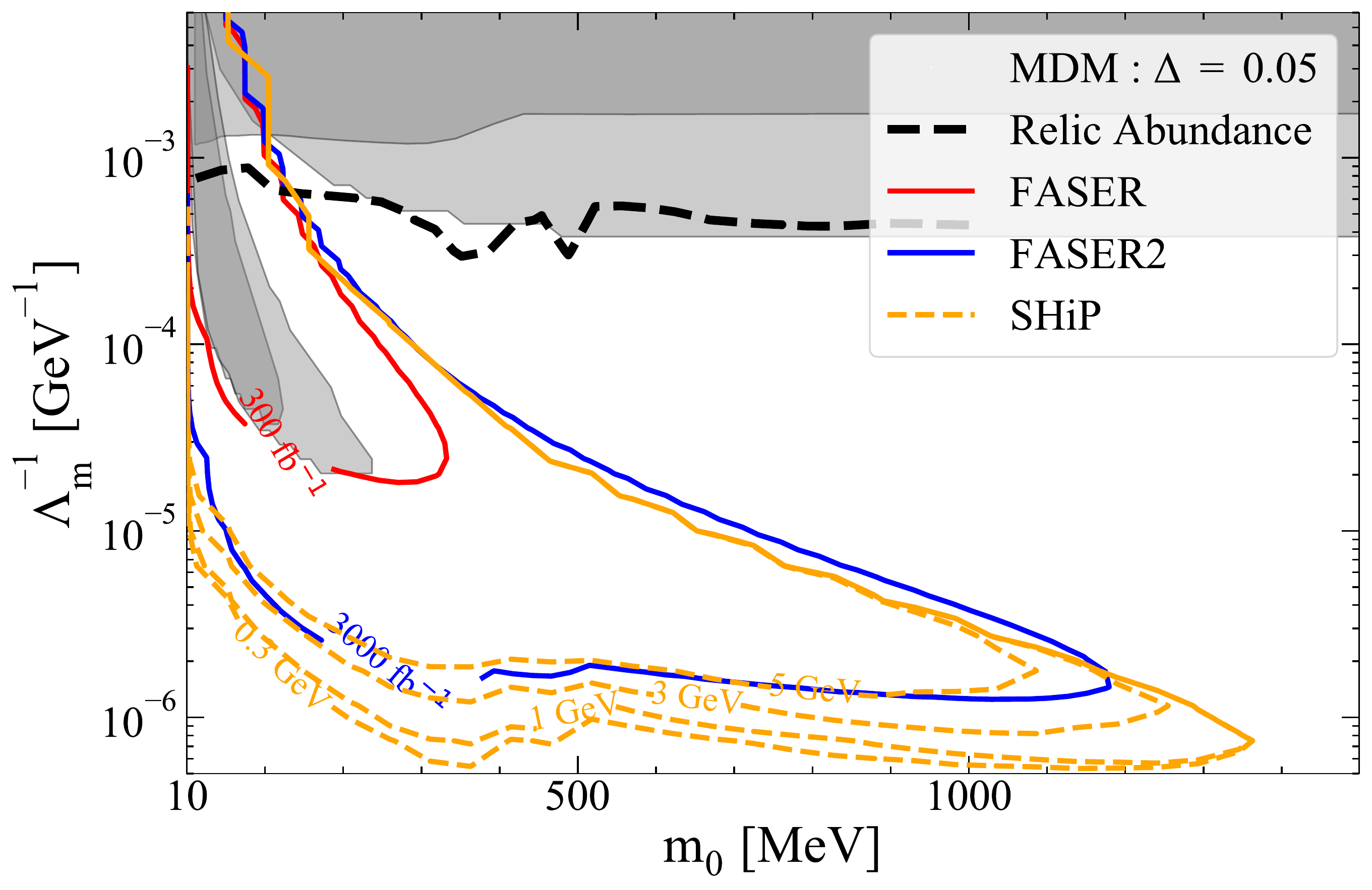}
    \includegraphics[width=0.49\textwidth]{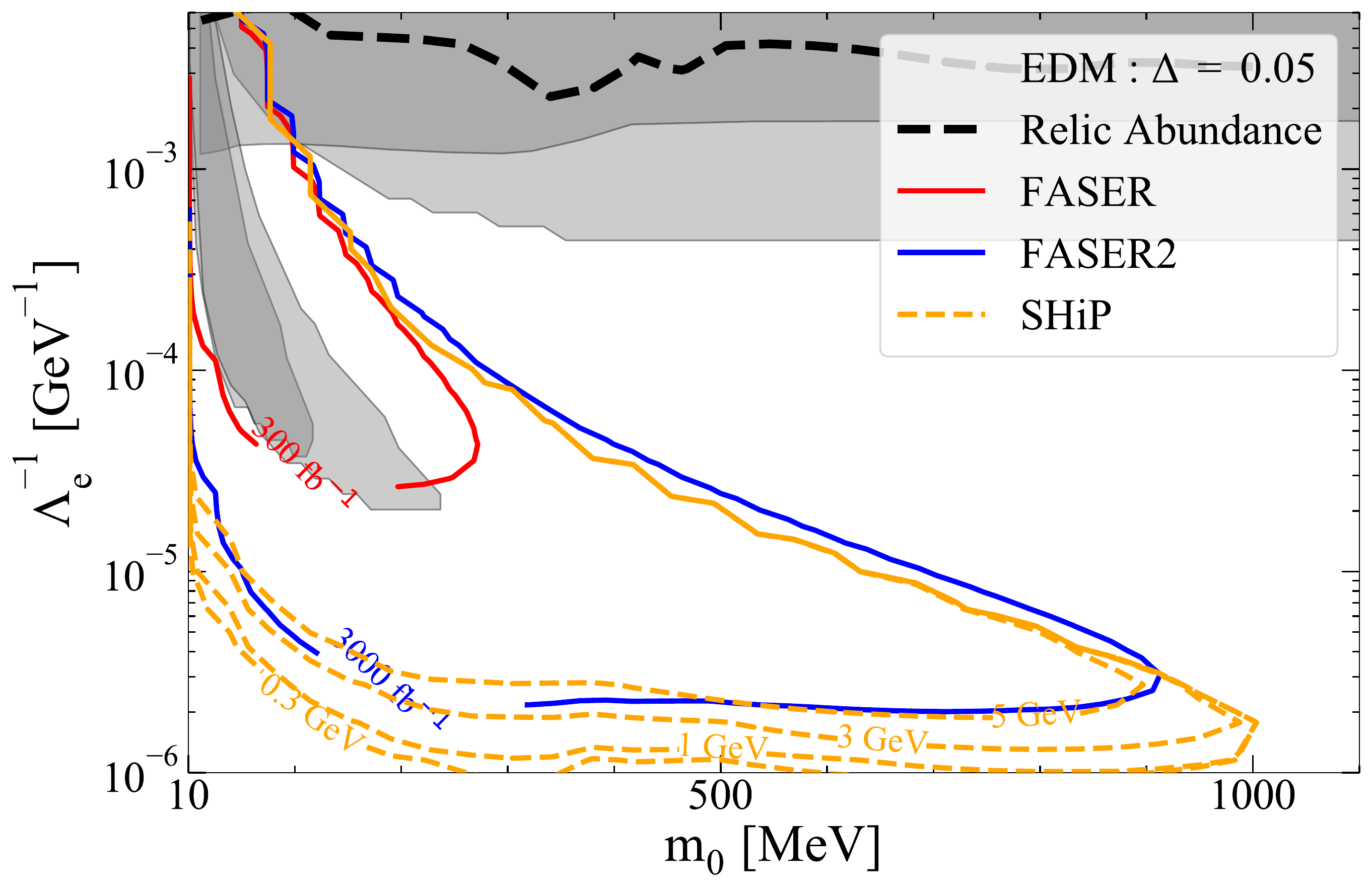}
    \caption{Similar to Fig.~\ref{fig:resultsplots}, but including contours representing the 
discovery reach of SHiP (dashed orange curves) for different monophoton energy 
thresholds.  Each such contour corresponds to the observation of three signal events 
for a total of $2\times 10^{20}$ protons on target.  Contours are shown for the 
thresholds $E_\gamma \geq 0.3$~GeV, $E_\gamma \geq 1$~GeV, $E_\gamma \geq 3$~GeV, 
and $E_\gamma \geq 5$~GeV.  We observe that SHiP has a comparable reach to FASER2
for large $\Delta$, even for larger detection thresholds, but has far less 
sensitivity than FASER2 for small $\Delta$, even for a detection threshold 
$E_\gamma \geq 0.3$~GeV. 
\label{fig:shipplot}}
\end{figure*}

As described in Ref.~\cite{Ahdida:2704147}, SHiP is proposed to consist 
of a decay volume shaped as a 50-meter-long pyramidal frustum with an upstream 
face 45 meters from the target with dimensions 5~m $\times$ 2~m and a downstream 
face with dimensions $11~{\rm m}\times 6~{\rm m}$.  The SHiP detector is 
expected to be able to detect photons with energies 
$E_\gamma \gtrsim 300$~MeV~\cite{Golutvin:2022}.  However, 
resolving the monophoton signal that results from $\chi_1\rightarrow \chi_0\gamma$ 
decay from background may require a somewhat higher photon-energy threshold.
For example, in assessing the prospects for detecting a monophoton signature 
of dark-photon decay in the context of a dark-axion-portal model,
the authors of Ref.~\cite{deNiverville:2019xsx} adopted a threshold 
$E_{\gamma}>2$~GeV based on comparisons with the process in which the
decay of the dark photon decay yields a $e^+e^-$ pair rather than a photon.
However, until a detailed monophoton study for SHiP is performed, it remains 
unclear precisely what threshold should be adopted in searches of this sort.
Thus, in what follows, we consider a range of possible choices for this 
threshold.

The procedure we use for determining the expected number of signal events at 
SHiP for any given combination of $m_0$, $\Delta$, and $\Lamcalo$ is analogous 
to the procedure described in \secref{fixed target} that we use in determining 
the expected number of events at other fixed-target experiments.
We calculate the differential cross-section for $\pi^0$ production due to the 
scattering of protons in the beam with a molybdenum target using the {\tt{BdNMC}} 
code package.  We then scale this cross-section by $\mathcal{N}_M^{({\rm POT})}$ as
in Eq.~(\ref{eq:RescaleDiffXSecMeson}) to obtain differential production 
cross-sections for heavier neutral mesons.  The $\mathcal{N}_M^{({\rm POT})}$ for
all relevant meson species except the vector meson $\psi(2S)$ were provided in 
Ref.~\cite{Chu:2020ysb}.  We calculate the number of $\psi(2S)$ mesons produced 
per proton using \texttt{Pythia 8.2} and find that
$\mathcal{N}_{\psi(2S)}^{({\rm POT})} = 10^{-6}$. 
We then determine the expected 
number of $\chi_1$ decays per proton on target within the decay volume of
the SHiP detector from the branching fractions given in Appendix~\ref{app:BFmesons},
assuming the detector geometry described above. 
 
In Fig.~\ref{fig:shipplot}, we show contours (orange dashed curves) of the 
discovery reach of the SHiP experiment within our parameter-space region of 
interest for $2\times 10^{20}$ total protons on target --- the number 
expected over roughly 10 years of operation~\cite{SHiP:2015vad}.
Since this discovery reach depends sensitively on the threshold
energy for a single photon, we plot contours for several different values of
this threshold: $E_\gamma \geq 0.3$~GeV, 1 GeV, 3 GeV, and 5 GeV.  As in 
Fig.~\ref{fig:resultsplots}, the panels shown in the left and
right columns correspond to the choice of MDM and
EDM operators, respectively, while the panels shown in different rows
correspond to different values of $\Delta$.  Contours indicating the reach of
FASER with an integrated luminosity of $300$~fb$^{-1}$ (red curves) and the
reach of FASER2 with $3000$~fb$^{-1}$ (blue curves) are also shown for 
comparison.

We observe from Fig.~\ref{fig:shipplot} that for small values of $\Delta$, 
the SHiP detector has little sensitivity within our parameter-space region
of interest.  This is primarily a reflection of the fact that for 
small mass splittings, the lab-frame energy of the photon produced by 
$\chi_1$ decay typically lies below the threshold for detection --- even if 
that threshold is as low as 300~MeV.
By contrast, for large values of $\Delta$, a larger fraction of the 
photons have energies that exceed the detection threshold
and the discovery reach of SHiP is far greater.  For example, for 
$\Delta = 0.05$, the reach of SHiP and FASER2 is very roughly similar for 
comparable running times.

\bibliography{references}

\end{document}